\documentclass[preprint,12pt]{aastex}

\newcommand{\etal}{{\em et al.\ }}

\def\ion[#1 #2]{#1\,{\sc #2}}
\def\lamb[#1]{#1\,{\AA}}
\def\lambr[#1-#2]{{{#1}--{#2}\,{\AA}}}
\def\rat[#1 #2]{#1/#2}
\def\serts89{SERTS-89}

\def\tabul{\hbox{\raise 0.75pt\hbox{$\triangleleft$}}}
\def\ergs[#1]{#1 {ergs}~{cm$^{-2}$}\,{s$^{-1}$}\,{sr$^{-1}$}}
\def\dens[#1]{10$^{#1}$\hskip 1.5pt{cm$^{-3}$}}
\def\densr[#1 #2]{10$^{#1}$\hskip 1pt{--}\hskip .5pt{10$^{#2}$}\hskip 1.5pt{cm$^{-3}$}}
\def\fl[#1 #2]{{#1}$\pm${#2}}
\def\orb[#1 #2]{{$#1^{#2}$}}
\def\ls[#1 #2]{{$^{#1}${#2}}}
\def\tm[#1 #2 #3]{{$^{#1}${#2}$_{#3}$}}

\shortauthors{Landi, Hutton, Brage, Li}
\shorttitle{Hinode/EIS measurements of active region magnetic fields}

\begin{document}

\title{Hinode/EIS measurements of active region magnetic fields}

\author{E. Landi}
\affil{Department of Climate and Space Sciences and Engineering, University of Michigan}

\author{R. Hutton}
\affil{Institute of Modern Physics, Fudan University, Shanghai 200433, People's Republic of China}

\author{T. Brage}
\affil{Department of Physics, Lund University, Box 118, SE-22100 Lund, Sweden}

\author{W. Li}
\affil{Department of Physics, Lund University, Box 118, SE-22100 Lund, Sweden}
\affil{Department of Materials Science and Applied Mathematics, Malm\"o University, SE-20506, Malm\"o, Sweden}

\begin{abstract}
The present work illustrates the potential of a new diagnostic technique that allows the
measurement of the coronal magnetic field strength in solar active regions utilizing a handful 
of bright \ion[Fe x] and \ion[Fe xi] lines commonly observed by the Hinode/EIS high-resolution
spectrometer. The importance of this new diagnostic technique lies in two basic facts: 1) the 
coronal magnetic field is probably the most important quantity in coronal physics, as it 
is at the heart of the processes regulating Space Weather and the properties of the solar 
corona, and 2) this technique can be applied to the existing EIS archive spanning from 2007 
to 2020, including more than one full solar cycle and covering a large number of active 
regions, flares, and even coronal mass ejections. This new diagnostic technique opens the 
door to a whole new field of studies, complementing the magnetic field measurements from 
the upcoming DKIST and UCoMP ground based observatories, and extending our reach to active 
regions observed on the disk and until now only sampled by radio measurements. In this work 
we present a few examples of the application of this technique to EIS observations taken at 
different times during the EIS mission, discuss its current limitations and the steps to 
improve its accuracy. We also present a list of EIS observing sequences whose data include 
all the lines necessary for the application of this diagnostic technique, to help the solar 
community navigate the immense set of EIS data and to find observations suitable to measure 
the coronal magnetic field.
\end{abstract}

\keywords{Sun: corona -- Sun: magnetic field}

\section{Introduction}

The magnetic field of the solar corona is one of the most critical parameters in solar physics,
as it lies at the core of most manifestations of coronal physics and of the interactions between
the Sun and its planetary system. In fact, the magnetic field confines and structures coronal
plasmas at all temperature and spatial scales, from bright points (few tens of arcseconds in size), 
to coronal loops and active regions (arcminutes) and streamers (solar radii). Also, regions of 
open and closed magnetic fields have very different properties, with the former being the site
of the acceleration of the fast solar wind and possibly of a part of the slow solar wind (Stakhiv 
\etal 2015, 2016).  Several types of magnetic waves have been proposed as candidates to heat the 
solar coronal plasmas up to multimillion degrees temperatures, as well as to accelerate the solar 
wind along open magnetic field lines (e.g. Cranmer 2009, Gombosi \etal 2018 and references therein). 
Magnetic reconnection has been suggested as a viable alternative for coronal heating and wind 
acceleration (Cranmer 2009), and is thought to be the trigger of both flares and coronal mass 
ejections (Webb \& Howard 2012 and references therein).

Despite its importance, the coronal magnetic field has proved elusive to infer directly, 
and very few measurements have been carried out. The photospheric magnetic field has been routinely
measured by a number of space missions in the last few decades by instruments such as SoHO/MDI
(Scherrer \etal 1995), Hinode/SOT (Suematsu \etal 2008) and SDO/HMI (Scherrer \etal 2012), and 
it is a fundamental component for local and global models of the solar atmosphere. Measurements 
of the coronal magnetic field are more difficult, because of the weakness of its signatures, 
and of the complexity of the measurements. Indirect measurements have been carried out studying 
coronal loop seismology (De Moortel \etal 2016) and magnetic field morphology has been empirically 
derived by active region plasma distribution. The only direct measurements so far have been 
obtained with radio observations, and spectropolarimetry of visible and near infrared (NIR) 
radiation, mostly with the Coronal Magnetometer and Polarimeter (CoMP, Tomczyk \etal 2008). 

To improve on CoMP measurements, and to provide at the same time measurements both of the 
magnetic field and of coronal plasma properties, two new ground based instruments are being 
built and are becoming operational in 2020: the Upgraded CoMP (UCoMP, Tomczyk \etal 2020, 
in preparation), and most importantly the Danel K. Inohue Solar Telescope (DKIST). Both 
instruments will measure the coronal magnetic field through visible and NIR spectropolarimetry, 
with a vastly different field of view and resolution: while the DKIST observatory (which will 
comprise a host of different instruments) will provide high spatial resolution measurements 
over a small field of view, UCoMP will provide near-simultaneous lower resolution measurements 
over the entire solar corona.

However, ground based spectropolarimetric magnetic field measurements of the coronal magnetic
fields suffer from a few fundamental limitations. First, they are affected by the day/night
cycle and by atmospheric transmission and scattering, although such a limitation has been 
overcome by space-based instruments. Second, since they 
observe in the visible and NIR wavelength ranges, they can only carry out measurements 
at the limb using coronographs, as the photospheric brightness prevents any measurements 
of coronal emission to be carried out on the disk. Third, limb observations in the visible
and NIR can only provide magnetic field orientation in the plane of the sky (through linear 
polarization -- Stokes U and Q) and the magnetic field component along the line of sight 
(through circular polarization -- Stokes V), so that the total magnitude of the magnetic 
field can not be reconstructed. Furthermore, the signal of Stokes V is weaker than the 
signal of Stokes U and Q, so that often only the magnetic field orientation in the plane of 
the sky can be measured.

The only measurements of coronal magnetic fields on the disk are done using radio measurements, 
sometimes in combination with Extreme Ultraviolet (EUV) observations. These measurements, however, 
are typically able to yield magnetic field strengths usually in excess of $\approx$200~G, and 
thus are most suitable for active regions. For example, Brosius \etal (2002) and Brosius \& 
White (2006) combined EUV and radio observations of active regions on the disk to infer magnetic 
field strengths in the 580-1750~G range. White \etal (2002) measured the magnetic field of a 
flaring loop and found that it ranged from $\approx$800~G at the footpoints to 224~G at 
the top; similar values were found by Nindos \etal (2000) (870~G to 280~G). Still, radio 
measurements are able to provide the magnetic field at different heights within the same 
magnetic structure, but are not able to determine the actual height of the structure itself, 
thus leaving a signficant uncertainty on the 3D orientation of the magnetic field vector.

Disk observations of the solar corona have been routinely carried out by imaging instruments
and high resolution spectrometers observing in the X-ray, EUV and 
Ultraviolet (UV) wavelength ranges by a host of rocket flights and space missions (e.g. 
Del Zanna \& Mason 2018 and references therein). These instruments constitute the backbone 
of our studies of the solar corona, as these wavelength ranges include a wealth of spectral 
lines and continuum radiation that provide plasma diagnostic tools allowing us to measure 
fundamental plasma properties such as electron densities and temperatures, plasma motions 
and elemental composition (Phillips \etal 2008, Del Zanna \& Mason 2018). However,
the effects of coronal magnetic fields on the intensities, wavelengths and profiles of 
emission lines in the X-ray, EUV and UV ranges are too small to be detected, so that no 
diagnostic tool is available to measure the magnetic field directly from line intensities 
in these wavelength ranges. 

Recently, Li \etal (2015, 2016) and Si \etal (2020a) reported on a peculiar atomic physics
configuration which makes the wavefunction composition of a low-energy metastable atomic 
level of \ion[Fe x] sensitive to the presence of an external magnetic field. This sensitivity 
would remain a curious feature of a single, metastable \ion[Fe x], if this level didn't 
happen to significantly contribute to the emission of one of the strongest lines in the 
\ion[Fe x] spectrum in the solar corona 
at 257.26~\AA. The properties of this metastable level cause the intensity of the line 
it emits to be directly and significantly affected by the local magnetic field strength;
since this ion is usually formed in the corona, this property opens a window through which
we can directly measure the magnetic field strength in the solar corona, which was first 
explored by Si \etal (2020b). 

The 257.26~\AA\ line has been routinely observed by the EIS spectrometer (Culhane \etal 2007)
on board the Hinode satellite (Kosugi \etal 2007), along with many other \ion[Fe x] lines for 
more than one solar cycle (2007 to mid-2020 at the time of writing). The vast data set 
accumulated by the EIS mission can be utilized to measure the coronal magnetic field strengths
throughout the solar cycle. The first application of this line to determine magnetic fields in a
Hinode/EIS active region has been described by Si \etal (2020b).

The goal of this paper is to refine and extend the diagnostic technique by making use of a few 
of the strongest \ion[Fe x] and \ion[Fe xi] lines routinely observed by the Hinode/EIS spectrometer 
from 2007 to 2020 to measure the magnetic field of the solar corona, present some examples, and 
discuss this diagnostic technique's strengths and limitations, as well as its synergies with the 
upcoming DKIST and UCoMP observatories. Future papers will apply this diagnostic technique to 
a number of different open problems in the solar corona. Section~\ref{method} introduces the 
diagnostic technique and its physical bases, Section~\ref{observations} describes the Hinode/EIS 
observing sequences that can be used to apply the technique, as well as the actual observations 
we analyzed, and Section~\ref{results} presents the magnetic field measurements we obtained. 
Section~\ref{uncertainties} discusses the uncertainties in the present method,
and Section~\ref{discussion} suggests future steps to improve on this technique.

\newpage

\section{Methodology}
\label{method}

\subsection{History of Magnetically Induced Transitions}

The influence of magnetic fields on atomic energy levels has a history going back to Zeeman 
in 1896. The Zeeman effect is used in many areas to measure the strength of external magnetic 
fields. What is much less known is the influence of external magnetic fields on the lifetimes 
of long lived levels, through an introduction of a new decay channel -- potentially resulting 
in new spectral features. The idea that a magnetic field could shorten the lifetime of a long 
lived level was investigated in the 1960s and was labelled {\em Zeeman quenching} by Feldman 
\etal (1967). The quenching was only considered through the shortening of, for example, the 
\orb[1s2s2p ] \tm[4 P 5/2] lifetime in \ion[Li i] and as an error source when trying to 
estimate the lifetimes in expected field-free space. The idea of a new or a change of an existing 
feature due to an external field did not seem to have any practical applications and the 
studies did not flourish. More recently, Beiersdorfer \etal (2003) used the Electron Beam Ion 
Trap (EBIT) at the Lawrence Livermore National Laboratory for the first observation of a spectral 
line induced by an external magnetic field -- the \orb[2p 5]\orb[3s ] \tm[3 P 0]-\orb[2p 6] \tm[1 S 0] 
transition in Ne-like Argon. This transition is strictly forbidden (being a J=0 to J=0) in 
the absence of an external magnetic field (or nuclear spin). The induction of this transition 
requires a field of a few Tesla and was considered mainly of importance for understanding the 
atomic structure (as well as predicting nuclear properties in non-zero spin isotopes), not 
for applications to astrophysical plasmas: the required field and low density did not offer 
any possible observations. 

Some years later a study was initiated by some of the present authors (TB, RH, WL) 
to find transitions induced by much lower magnetic field 
strengths. The basic atomic structure needed for the induction of a spectral line through 
an external magnetic field is two energy levels, relatively close in energy, where one has 
an allowed decay, and therefore short lifetime, while the other one should be at least 
metastable with a considerably longer lifetime. If the effect should be observable for 
small magnetic fields, the energy splitting between the levels has to be small, so the search 
focused on 'accidental' pseudo-degeneracy, induced by level crossing in the gross 
structure of ions, along an isoelectronic sequence. It represented a major breakthrough 
when the project focused on the interesting isoelectronic behavior of the 
\orb[3s 2]\orb[3p 4]\orb[3d ] \tm[4 D ] term in Cl-like ion, observed by 
isoelectronic studies and observations as reported in the NIST database (Kramida \etal 2019). 

The \tm[4 D ] term has four energy levels of which the \tm[4 D 7/2] level is metastable and 
can only decay to the ground J=3/2 level of the \orb[3s 2]\orb[3p 5] \tm[2 P ] term
by a forbidden, 
magnetic quadrupole (M2), in the absence of an external field, since J must change by 2 units. 
However, the \tm[4 D 5/2] level has an allowed, electric dipole (E1) decay channel to the 
\tm[2 P 3/2] level (albeit spin-changing), leading to a lifetime five order of magnitudes 
shorter. The ordering of these two levels changes along the sequence and as discussed by 
Li \etal (2015) the minimum energy separation between these two levels occurs for \ion[Fe x], 
leading to a pseudo-degeneracy. In this work it was shown for the first time that there was 
definitely a magnetic induced transition, MIT, from the \tm[4 D 7/2] level of \ion[Fe x], 
induced by the external magnetic field mixing of the two levels -- the J=7/2 acquires some 
of the J=5/2 levels properties, including a decay channel to the ground term. It was also 
clear that this MIT was sensitive to fairly small magnetic fields. One problem discussed 
by Li \etal (2015) was the fact that the energy splitting of the two levels, which is 
crucial for the required sensitivity of the MIT to the low enough external magnetic field
strengths, was not known accurately enough. In a following work by Li \etal 
(2016), the Shanghai high temperature superconducting EBIT with known magnetic field
was used to obtain a value for the \tm[4 D 5/2]-\tm[4 D 7/2] fine structure energy. 
The energy obtained was 3.5~cm$^{-1}$, which agreed with the astrophysical estimate of 
0 -- 5~cm$^{-1}$, however with a large uncertainty (the astrophysical estimate do not 
offer any explicit guidance on uncertainty). Li \etal (2016) also showed that a line 
ratio of the \tm[4 D 7/2] and \tm[4 D 5/2] blended transitions with the transition from 
one of the other \tm[4 D ] levels could be used as a magnetic field diagnostic. 
Inspired by these initial efforts, Judge \etal (2016) used spectra of the solar corona 
from SkyLab to improve on the uncertainty of the \tm[4 D 5/2,7/2] fine structure energy 
and arrived at the result of 3.6$\pm$2.7~cm$^{-1}$. Recently this fine structure value 
was used in a determination of the magnetic field for an active region of the solar 
corona from spectral data from Hinode (Si \etal 2020b). Even more recently, Landi \etal
(2020) improved on the measurement by Judge \etal (2016) using spectra from the SOHO/SUMER
high resolution spectrometer, to
obtain an energy separation of 2.29$\pm$0.50~cm$^{-1}$, which significantly lower
the uncertainty (see Section~\ref{uncertainties}); in the present work, we will be 
using this value.

\subsection{Conceptual description of the MIT}

The core of the method discussed in this paper is the concept of mixing of atomic states 
of the same parity but different J-values. The strength of the mixing depends strongly on 
the separation of the levels in energy and the magnetic field strength (both squared). 
All interactions between the electrons are diagonal in the total angular momentum quantum 
number J. Therefore two levels such as the \orb[3s 2]\orb[3p 4]\orb[3d ] \tm[4 D 5/2] and 
\tm[4 D 7/2] in Cl-like ions cannot be mixed due to electron-electron interaction, even 
if the levels are basically energy degenerate. To induce this mixing and thereby the decay 
of the metastable level, we need to introduce interactions outside the electronic cloud, 
e.g. an externally applied magnetic field, which can mix levels with J differing by 0 or 
$\pm$1. The other possibility for this mixing to occur, is through a nuclear 
spin, which changes the total angular momentum of the ion. In the cases reported here, we only 
consider nuclei with zero spin. 

It is important to note that the mixing of the \orb[3s 2]\orb[3p 4]\orb[3d ] \tm[4 D 5/2] 
and \tm[4 D 7/2] and the resulting MIT in Cl-like Fe can be induced by unexpectedly small 
fields, of the order of a few hundred Gauss (or less). This is to be compared to the 
internal magnetic field of \ion[Fe x], caused by the orbiting electrons and their spins, 
which is of the order of many hundreds or even thousands of Tesla. As a metaphore, one 
could use that the external field only tickles the ion, but the accidental pseudo-degeneracy 
of two levels causes it to decay with a EUV-photon from the metastable state.

\subsection{Measuring the magnetic field with the MIT}

The most direct way to utilize the magnetically induced transition to measure the ambient
magnetic field is to determine the MIT/M2 branching ratio from the observations. The first
step is to disentangle the intensity of the two spectral lines at 257.26~\AA\ in order
to separate the E1 component from the MIT+M2 component. However, the energy separation between 
the \tm[4 D 7/2] and \tm[4 D 5/2] levels gives rise to a difference in the corresponding 
wavelengths of only $\approx$4~m\AA, which is by far too small to be resolved by either 
Hinode/EIS or any other high-resolution spectrometer built so far, and is also much 
smaller than the line broadening in the corona. Thus, the MIT+M2 fraction 
needs to be determined by indirect means. 

In principle, any \ion[Fe x] line intensity ratio involving the 257.26~\AA\ is dependent 
on the magnetic field. However, since the metastable nature of the upper \tm[4 D 7/2] level 
causes all ratios involving this line to also be density sensitive, an independent measurement 
of the electron density needs to be available before determining the magnetic field.

Si \etal (2020b) proposed the use of two \ion[Fe x] spectral lines to measure the magnetic 
field: 174.53~\AA, 175.26~\AA, in addition to the 257.26~\AA. Their technique relied on 
comparing the measured 174.53/257.26 intensity ratio with theoretical estimates carried
out at multiple values of the ambient magnetic field strength, $B$. The electron density 
was estimated using the 175.26/174.53 ratio, which has two advantages: first, it is the 
most density-sensitive ratio in the EUV \ion[Fe x] spectrum; second, these two lines are 
strong features in the spectra. However, they are located in wavelength at the edge of 
the EIS passband, leading to their signal-to-noise ratio (SNR) being low in EIS 
observations. Therefore they are seldom included in the line selection to be telemetered 
down from the satellite. This is especially true when images of the magnetic field in 
active regions are sought, for which rebinning can be limited and thus, even if available, 
the 174.53~\AA\ and 175.26~\AA\ may have too low SNR. Thus, while the approach in itself 
is powerful, it can only be applied to a limited number of EIS observations.

In this paper, we developed a different approach, which utilizes the brightest, and most
commonly observed \ion[Fe x] spectral line to obtain the same measurement: the 184.54~\AA\
line. Furthermore, we utilized the CHIANTI database (V 9.0, Dere \etal 1997, 2019) to 
disentangle both the MIT and M2 line intensities from the observations, to determine 
their branching ratios and compare the results directly to the atomic physics 
calculations as a function of $B$.

The presence of a MIT contribution can be determined by measuring the intensity $I_{MIT}$ of 
the 257.26~\AA\ line in excess from the value predicted by CHIANTI neglecting its presence, as

\begin{equation}
I_{MIT} = I_{257}-I_{184}\times R{\left({257/184}\right)}
\label{mit}
\end{equation}

\noindent
where $I_{257}$ and $I_{184}$ are the measured intensities for the 257.26~\AA\ amd 184.54~\AA\
lines, respectively, and $R{\left({257/184}\right)}$ is the intensity ratio predicted by 
CHIANTI including both the E1 and M2 components, but not the MIT one. That is, 
Equation~\ref{mit} measures the excess emission in the 257.26~\AA\ that CHIANTI can not 
account for with only the M2 and E1 transitions.

\begin{figure}[!t]
\begin{center}
\includegraphics[width=11.0cm,height=16.0cm,angle=90]{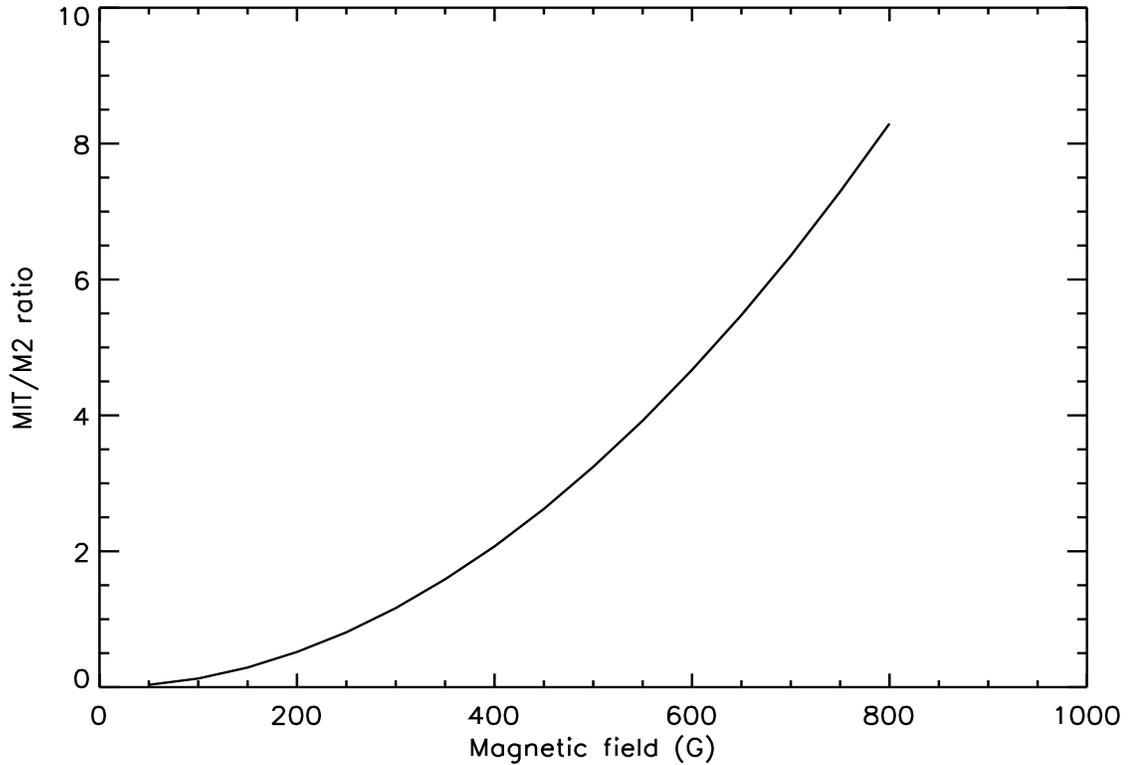}
\end{center}
\caption{$A_{MIT}/A_{M2}$ branching ratio as a function of magnetic field magnitude, in Gauss.}
\label{mit_m2_ratio}
\end{figure}

The intensity $I_{M2}$ of the M2 transition can be directly determined as

\begin{equation}
I_{M2} = I_{184}\times R{\left({M2/184}\right)}
\label{m2}
\end{equation}

\noindent
where $R{\left({M2/184}\right)}$ is the ratio between the M2 component and the 184.54~\AA\ 
line, also predicted by CHIANTI. The ratio $I_{MIT}/I_{M2}$ can then be directly compared 
with the branching ratio predicted by Li \etal (2015), and shown in Figure~\ref{mit_m2_ratio}.

There are two things to note. First, both the $R{\left({257/184}\right)}$ and the $R{\left({M2/184}\right)}$
are sensitive to the electron density $n_e$, the sensitivity being largest when $n_e>10^9$~cm$^{-3}$, 
typical of active regions. Thus, an independent estimate of the electron density is necessary for this 
ratio. Si \etal (2020b) chose the best line ratio to carry out the estimate, namely the 174.53/175.26 ratio.
This ratio should be used whenever possible. However, since Hinode/EIS observations have low SNR for 
174.53~\AA\ and 175.26~\AA\ lines, and often do not include them, it is often necessary to use another ratio. 

The other \ion[Fe x] lines in the EIS range are weaker, and while
a few individual observations can include them, they can not be used for extensive application to 
magnetic field measurements, and thus it is necessary to resort to density sensitive ratios from
other ions. These ions need to be formed close in temperature to \ion[Fe x], and need to provide
line pairs with strong lines routinely observed and telemetered down by EIS. The best candidate is
\ion[Fe xi] 182.17/(188.22+188.30) line ratio. These \ion[Fe xi] lines are strong, and extensively 
included in EIS observing sequences; the 188.2 doublet is only partially resolved by EIS, but its 
intensity can be easily measured either by a double Gaussian fit or simply by summing all the counts 
under the profile and subtracting the background. This doublet is sufficiently strong and isolated 
to allow for both options. These lines are stronger than those available for density diagnostics 
in the EIS line list for \ion[Fe ix] -- the other closest Fe ion -- or for ions of other elements.

In principle, the \ion[Fe viii] 185.21/186.60 intensity ratio can also be used as density diagnostics 
to measure the magnetic field in cold, isolated loop structures when no other lines are present. Both 
these lines are bright and isolated, are often telemetered to the ground, and provide a sensitive 
density diagnostic line pair. However, care should be taken in checking that the intensity distribution 
of \ion[Fe viii] and \ion[Fe x] is the same, to minimize errors and make sure that the emitting plasma 
is the same and the temperature difference between these two ions is not a problem. A similar caution
needs to be taken when using the other line pairs commonly included in EIS observing sequences: 
\ion[Fe xii] 186.8/195.1 and \ion[Fe xiii] 203.8/202.0: both ions are formed at a significantly 
higher temperature than \ion[Fe x]. We discuss the uncertainties due to electron density determination 
in Section~\ref{uncertainties}.

Another critical point to note is that the 257.26~\AA\ line is the only strong \ion[Fe x] line in the
long wavelength (LW) detector, while all others are observed in the short wavelength (SW) one. This makes
the relative calibration of the two detectors a critical factor in the measurement of coronal magnetic
fields. The EIS intensity calibration was measured before launch (Lang \etal 2006), but it has been 
subsequently revised to account for inaccuracies and sensitivity decrease with time. Unfortunately, 
two competing calibrations have been produced (Warren \etal 2014 and Del Zanna 2013 -- HPW and GDZ, 
respectively) which provide different relative SW/LW calibration factors, which in addition depend 
on time in a different way.  This provides a critical uncertainty to the present measurements which 
will also be discussed in Section~\ref{uncertainties}.

\section{Observations}
\label{observations}

\subsection{EIS observing sequences suitable for magnetic field diagnostics}

In order to test the magnetic field diagnostic technique, we produced magnetic field maps on several
active regions observed during the entire EIS mission. 
The results we report in this paper are just a drop in the ocean, as the EIS mission has developed 
plenty of observing sequences which include the necessary lines to carry out magnetic field measurements. 
In order to help users select suitable sequences to measure the coronal magnetic field, a list of these
sequences is reported in Tables~\ref{suitable_seq_1} to \ref{suitable_seq_3}. The vast majority of
them either includes the entire Hinode/EIS spectrum, or the four \ion[Fe x] and \ion[Fe xi] 
lines we used to carry out the present diagnostics. We also report a few more where \ion[Fe xi]
density diagnostics is not available, but either the \ion[Fe viii] 185.2/186.6 and/or the \ion[Fe xii]
186.8/195.1 ratios were available: these sequences may also be used for magnetic field diagnostics,
although care must be taken in ensuring that the plasma structures observed by \ion[Fe viii,x,xii] 
are the same. We hope that this list will be useful to the reader to identify data sets suitable for
the measurements of the magnetic field in the solar corona.

\begin{table}
\footnotesize
\begin{center}
\begin{tabular}{lrrccl}
Acronym                & FOV            & Exp.time      & Slit     & Full      & Notes \\
                       & (arcsec$^2$)   &          (s)  & (arcsec) & spectrum  &       \\
\hline
 & & & & \\
AKATSUKI\_01\_SI                   &  45$\times$512 &  240          & 1        & Y \\ %   38m28s
arm\_loop\_ne                      &  82$\times$400 &  25           & 2        &   \\ % 4h25m26s
AR\_velocity\_map                  & 330$\times$304 &  40           & 1        &   \\ % 5h15m23s 
AR\_velocity\_map\_v2              & 459$\times$384 &  40           & 1        &   \\ % 5h40m35s
AR\_spectral\_atlas\_3             & 120$\times$120 &  40           & 2        & Y \\ % 1h 9m14s
Atlas\_060x512\_45s                &  60$\times$512 &  45           & 1        & Y \\ %   49m35s
Atlas\_060x512\_60s                &  60$\times$512 &  60           & 1        & Y \\ % 1h 5m20s
Atlas\_30                          & 120$\times$160 &  30           & 2        & Y \\ %   33m50s
Atlas\_60                          & 120$\times$160 &  60           & 2        & Y \\ % 1h 5m20s
Atlas\_120                         & 120$\times$160 & 120           & 2        & Y \\ % 2h 8m20s
cam\_artb\_cds\_a\_lite            & 200$\times$400 &  10           & 2        &   \\ %   21m 9s  ** 
cam\_artb\_lite\_v2                &  40$\times$120 &  10           & 2        &   \\ %    4m32s  ***
cam\_ar\_limb\_lite\_v2            & 323$\times$376 &  45           & 2        &   \\ %   21m 9s  **
cam\_ar\_temp\_lite                & 359$\times$400 &  30           & 2        &   \\ % 1h40m46s  
cam\_qs\_2as\_velo                 &  20$\times$200 &  15           & 2        &   \\ %   59m25s
cam\_qs\_2as\_context              & 120$\times$360 &  30           & 2        &   \\ %   33m50s
cavity\_dem\_1                     & 302$\times$352 & 100           & 2        &   & \ion[Fe viii] \\ % 4h29m34s
CLASP-1\_60x512\_OBS\_             &  60$\times$512 &  60           & 2        &   \\ %   32m51s
CLASP-1\_60x512\_CAL\_             &  60$\times$512 &  60           & 1        &   \\ % 1h 5m20s
CompS\_NonMax\_120                 &  80$\times$512 & 120           & 2        &   \\ % 1h25m41s 
CompS\_NonMax\_90                  &  80$\times$512 & 120           & 2        &   \\ % 1h 4m41s  
CompS\_NonMax\_60                  &  80$\times$512 & 120           & 2        &   \\ %   43m41s
CompS\_NonMax\_30                  &  80$\times$512 & 120           & 2        &   \\ %   22m41s
COMSCI\_QS5                        &  10$\times$512 &   5           & 1        & Y \\ % 2h42m45s
Cool\_loop\_response               &  90$\times$144 &  20           & 2        &   & \ion[Fe viii] \\ % 2h52m37s
Coronal\_rain\_1as2pos             &  40$\times$256 &  10           & 1        &   & \ion[Fe viii] \\ %      54s
dhb\_atlas\_30x512                 &  60$\times$512 & 120           & 2        & Y \\ % 1h 4m21s
dhb\_atlas\_120m\_30               &  60$\times$160 & 120           & 2        & Y \\ % 1h 4m21s
DIAG\_40x180\_s0\_30s              &  40$\times$176 &  30           & 2        &   \\ %   26m35s
dob\_bp\_slit\_raster              & 120$\times$160 &  30           & 2        &   & \ion[Fe xii] \\ %    33m50s
DRW001\_HI\_BRT\_SCAN              &   5$\times$240 &  60           & 1        &   \\ % 1h 5m20s
DRW001\_HI\_BRT\_V2                &   5$\times$240 &  30           & 1        &   \\ %    3m10s
DRW001\_HI\_BRT\_V3                &  10$\times$240 &  30           & 1        &   \\ %    5m57s
%Eclipse\_raster\_1                 & 107$\times$176 &  40           & 1        &   \\ %   
Eclipse\_raster\_2                 & 180$\times$176 &  30           & 2        &   \\ %   50m34s
 & & & & \\
\hline
\end{tabular}
\end{center}
\caption{\label{suitable_seq_1} \small Hinode/EIS observing sequences suitable for coronal
magnetic field diagnostics. Sequences where density diagnostics from \ion[Fe xi] is unavailable
but is provided by \ion[Fe viii] and/or \ion[Fe xii] ratios are indicated in the far-right
column.}
\end{table}

\begin{table}
\footnotesize
\begin{center}
\begin{tabular}{lrrccl}
Acronym                & FOV            & Exp.time      & Slit     & Full      & Notes \\
                       & (arcsec$^2$)   &          (s)  & (arcsec) & spectrum  &       \\
\hline
 & & & & \\
EL\_FULL\_CCD\_RASTER              &  14$\times$512 & 300           & 2        & Y \\ %   37m21s
EL\_FULL\_CCD\_SINGLE              &   4$\times$512 & 300           & 2        & Y \\ % 1h12m14s
EL\_abund\_plume\_SUMER            &  46$\times$512 & 100           & 2        &   & \ion[Fe viii] \\ %    55m45s
EL\_loop\_plume\_SUMER             &  61$\times$512 & 100           & 2        &   \\ %   55m38s
el\_loop\_diagn\_sumer             &  61$\times$280 &  50           & 1        &   \\ %   55m45s
EL\_WHI\_CH\_small                 &  62$\times$120 & 100           & 2        &   \\ %   55m38s
EL\_WHI\_CH\_LIMB                  &   2$\times$512 &               & 2        & Y \\ %   33m35s
EL\_SUMER\_UVCS                    &   2$\times$400 & 150           & 2        & Y \\ % 1h 6m49s  
EUNIS\_EIS\_CrossCal18             & 122$\times$512 &  50           & 2        &   \\ %   55m45s
FILL001                            &   1$\times$256 &  30           & 2        & Y \\ %    4m47s   ***
FELDMAN\_QSCH\_ATLASv1             &  60$\times$304 & 120           & 2        & Y \\ % 1h 4m21s
FOCUS\_STUDY                       &  60$\times$512 &  90           & 1        &   & \ion[Fe xii] \\ % 1h36m50s
fullccd\_sns30                     &   2$\times$160 &  30           & 2        &   \\ %   33m17s   ***
fullccd\_scan\_m30                 & 120$\times$160 &  30           & 2        & Y \\ %   17m40s   **
fullccd\_scan\_l30                 & 120$\times$160 &  30           & 2        & Y \\ %    9m18s   ***
GDZ\_300x384\_S2S3\_40             & 300$\times$384 &  40           & 2        &   \\ % 1h14m23s
GDZ\_DENS\_20x240\_ARL1            &  20$\times$240 &  10           & 2        &   \\ %    2m27s   ***
GDZ\_DENS\_20x240\_ARL2            &  20$\times$280 &  15           & 2        &   \\ %    3m20s   ***
GDZ\_360x288\_AR\_CONT2            & 360$\times$288 &   6           & 2        &   & \ion[Fe xii] \\ 16m54s ***
GDZ\_PLUME1\_2\_300\_150           & 300$\times$512 & 150           & 2        &   \\ % 3h19m42s
GDZ\_PLUME1\_2\_300\_50s           & 300$\times$512 &  50           & 2        &   \\ % 1h 8m27s
gdz\_off\_limb1\_60                & 487$\times$512 &  60           & 2        &   \\ % 1h 9m25s
GDZ\_QS1\_60x512\_60s              &  60$\times$512 &  60           & 2        &   & \ion[Fe xii] \\ %   32m15s *
HH\_QS\_RAS\_N01                   &  60$\times$120 &  60           & 1        & Y \\ % 1h 5m20s
HH\_QS\_RAS\_H01                   &  41$\times$120 &  60           & 1        & Y \\ %   43m55s
HPW001\_FULLCCD\_RAST              & 128$\times$128 &  90           & 1        & Y \\ % 3h26m10s
HPW001\_FULLCCD\_V2                &  26$\times$456 &  90           & 1        & Y \\ % 3h26m10s
HPW008\_FULLCCD\_RAST              & 128$\times$128 &  25           & 1        & Y \\ % 1h 0m34s
HPW008\_FULLCCD\_V2                & 128$\times$256 &  25           & 1        & Y \\ % 1h 0m34s
HPW009\_FULLCCD\_SAS               &   1$\times$128 &  25           & 1        & Y \\ %   54m27s ***
HPW020\_VEL\_FULLs1                &   1$\times$512 & 120           & 1        & Y \\ %   45m10s ***
%HPW023\_FULLCCD\_300s              &   1$\times$512 & 300           & 1        & Y \\ %    5m39s ***
HPW023\_FULLCCD\_V2                &   1$\times$512 & 300           & 1        & Y \\ %    5m39s ***
HPW023\_FULLCCD\_V2s2              &   2$\times$512 & 300           & 2        & Y \\ %    5m39s ***
HPW023\_FULLCCD\_V3s2              &   2$\times$512 & 100           & 2        & Y \\ %    2m 9s ***
 & & & & \\
\hline
\end{tabular}
\end{center}
\caption{\label{suitable_seq_2} \small Hinode/EIS observing sequences suitable for coronal
magnetic field diagnostics. Sequences where density diagnostics from \ion[Fe xi] is unavailable
but is provided by \ion[Fe viii] and/or \ion[Fe xii] ratios are indicated in the far-right
column.}
\end{table}

\begin{table}
\footnotesize
\begin{center}
\begin{tabular}{lrrccl}
Acronym                & FOV            & Exp.time      & Slit     & Full      & Notes \\
                       & (arcsec$^2$)   &          (s)  & (arcsec) & spectrum  &       \\
\hline
 & & & & \\
iiap\_ch\_sns\_v2                  &   2$\times$512 &  60           & 2        &   & \ion[Fe viii,xii] \\ % 3h15m16s
iiap\_ch\_ctxt\_v2                 & 120$\times$512 &  60           & 2        &   \\ % 1h 5m20s
KM\_FULLCCD\_128x256\_1slit\_80sec & 128$\times$256 &  80           & 1        & Y \\ % 3h 3m46s
KM\_FULLCCD\_128x512\_1slit\_80sec & 128$\times$512 &  80           & 1        & Y \\ % 3h 3m46s
kpd\_01\_qs\_60s                   &  56$\times$512 &  60           & 1        &   & \ion[Fe viii,xii] \\ % 1h 1m 1s
LANDI\_SCAN\_CH                    &  60$\times$512 & 240           & 2        &   \\ % 2h 7m21s
Large\_CH\_Map                     & 180$\times$512 &  60           & 2        &   & \ion[Fe viii,xii] \\ % 1h 5m20s
madj\_ech                          & 120$\times$512 &  60           & 2        &   & \ion[Fe viii,xii] \\ % 1h 5m20s
madj\_ech\_small                   &  24$\times$512 &  60           & 2        &   & \ion[Fe viii,xii] \\ %   13m22s **
prom\_rast\_v1                     &  81$\times$128 &  50           & 1        &   & \ion[Fe viii,xii] \\ % 1h13m54s
prom\_rast\_small\_v2              &   4$\times$128 &  25           & 1        &   & \ion[Fe viii,xii] \\ %   28m35s *
PRY\_CH\_density                   &  70$\times$200 & 100           & 2        &   & \ion[Fe viii,xii] \\ % 1h 2m46s
PRY\_footpoints\_v2                & 100$\times$140 &  25           & 2        &   \\ %   27m 2s *
PRY\_footpoints\_HI                & 180$\times$512 &  25           & 2        &   \\ %   42m42s
PRY\_footpoints\_HI2               & 180$\times$512 &  25           & 2        &   \\ %   42m42s
PRY\_loop\_footpoints              & 100$\times$216 &  30           & 2        &   \\ %   26m21s *
QS\_atlas\_offlimb                 & 220$\times$512 &  60           & 1        &   \\ %   47m18s
RED\_SUM\_EIS\_RAST                &  70$\times$200 &  90           & 1        &   \\ %   56m39s
RED\_SUM\_EIS\_SNS\_OL             &  70$\times$200 &  90           & 1        &   \\ %   56m39s
SI001FullRast                      & 256$\times$256 &  50           & 1        & Y \\ % 3h52m46s
SI002\_HiCadence\_AR               & 280$\times$512 &  50           & 1        & Y \\ %   27m56s *
SI\_Mercury\_slit                  &   2$\times$256 &  60           & 2        & Y \\ %    1m27s ***
SI\_Mercury\_slit\_v2              &   2$\times$256 &  20           & 2        & Y \\ %      45s ***
SI\_Venus\_slit                    &   2$\times$256 & 300           & 2        & Y \\ %    5m39s ***
SI\_Venus\_slit\_v2                &   2$\times$256 & 100           & 2        & Y \\ %    2m 9s ***
%SK\_DEEP\_5x512\_SLIT1             &   5$\times$512 & 120           & 1        &   & No dens diag \\
SYNOP001                           &   1$\times$256 &  30           & 1        & Y \\ %      30s **
SYNOP002                           & 128$\times$184 &  90           & 1        & Y \\ % 3h26m10s     
SYNOP003                           & 300$\times$96  &  45           & 2        & Y \\ % 2h 3m25s
SYNOP004\_400x400\_30              & 400$\times$400 &  30           & 2        &   \\ % 3h36m55s
Utz\_quiet                         &  22$\times$160 &  60           & 1        &   & \ion[Fe xii] \\ %    12m17s **
 & & & & \\
\hline
\end{tabular}
\end{center}
\caption{\label{suitable_seq_3} \small Hinode/EIS observing sequences suitable for coronal
magnetic field diagnostics. Sequences where density diagnostics from \ion[Fe xi] is unavailable
but is provided by \ion[Fe viii] and/or \ion[Fe xii] ratios are indicated in the far-right
column.}
\end{table}

\subsection{Data used in the present work}

For the present work, we only selected a few observations, requiring that their fields of view 
include an active region, and they include all four \ion[Fe x] and \ion[Fe xi] lines to carry 
out the present diagnostics. We also selected a few more to check the effects of using \ion[Fe xi] 
density and detector degradation effects later in the mission. These observations are listed in 
Table~\ref{datasets}.  These images were cleaned and calibrated using the standard EIS software. 
In addition, the detector slant was corrected so that the images obtained in the LW and SW were 
coaligned. 

\begin{table}
\tiny
\begin{center}
\begin{tabular}{ccccclc}
Date & Time & FOV              & Exp.time     & Slit          & Sequence    & Full      \\
     &      &     (arcsec$^2$) &          (s) &      (arcsec) &             & spectrum  \\
\hline
 & & & & \\
\multicolumn{7}{c}{Density comparison} \\
 & & & & \\
2007-Jun-02 & 19:56:12 & 128$\times$128 & 25   & 1" & HPW008\_FULLCCD\_RAST & Y \\
2010-Jun-21 & 14:24:01 & 120$\times$160 & 60   & 2" & ATLAS\_60             & Y \\
 & & & & \\
\multicolumn{7}{c}{Magnetic field imaging} \\
 & & & & \\
2007-Dec-10 & 00:19:27 & 459$\times$384 & 40   & 1" & AR\_velocity\_map\_v2 & N \\
2007-Dec-11 & 10:25:42 & 459$\times$384 & 40   & 1" & AR\_velocity\_map\_v2 & N \\
2007-Dec-12 & 03:26:43 & 459$\times$384 & 40   & 1" & AR\_velocity\_map\_v2 & N \\
2007-Dec-12 & 11:43:36 & 459$\times$384 & 40   & 1" & AR\_velocity\_map\_v2 & N \\
2007-Dec-13 & 12:18:42 & 459$\times$384 & 40   & 1" & AR\_velocity\_map\_v2 & N \\
2007-Dec-15 & 00:13:49 & 459$\times$384 & 40   & 1" & AR\_velocity\_map\_v2 & N \\
2007-Dec-15 & 18:15:49 & 459$\times$384 & 40   & 1" & AR\_velocity\_map\_v2 & N \\
2007-Dec-18 & 00:10:49 & 459$\times$384 & 40   & 1" & AR\_velocity\_map\_v2 & N \\
2007-Dec-18 & 18:13:41 & 459$\times$384 & 40   & 1" & AR\_velocity\_map\_v2 & N \\
 & & & & \\
\multicolumn{7}{c}{Time variation magnetic field} \\
 & & & & \\
2008-Jan-10 & 18:07:32 & 180$\times$512 & 25   & 2" & PRY\_footpoints\_HI   & N \\
2008-Jan-10 & 22:51:03 & 180$\times$512 & 25   & 2" & PRY\_footpoints\_HI   & N \\
2008-Jan-11 & 00:16:33 & 180$\times$512 & 25   & 2" & PRY\_footpoints\_HI   & N \\
2008-Jan-11 & 00:57:03 & 180$\times$512 & 25   & 2" & PRY\_footpoints\_HI   & N \\
2008-Jan-11 & 01:37:32 & 180$\times$512 & 25   & 2" & PRY\_footpoints\_HI   & N \\
2008-Jan-11 & 02:18:02 & 180$\times$512 & 25   & 2" & PRY\_footpoints\_HI   & N \\
2008-Jan-11 & 03:39:02 & 180$\times$512 & 25   & 2" & PRY\_footpoints\_HI   & N \\
2008-Jan-11 & 04:31:24 & 180$\times$512 & 25   & 2" & PRY\_footpoints\_HI   & N \\
2008-Jan-11 & 05:30:47 & 180$\times$512 & 25   & 2" & PRY\_footpoints\_HI   & N \\
2008-Jan-11 & 06:11:17 & 180$\times$512 & 25   & 2" & PRY\_footpoints\_HI   & N \\
2008-Jan-11 & 07:11:06 & 180$\times$512 & 25   & 2" & PRY\_footpoints\_HI   & N \\
2008-Jan-12 & 13:20:33 & 180$\times$512 & 25   & 2" & PRY\_footpoints\_HI   & N \\
2008-Jan-12 & 14:01:04 & 180$\times$512 & 25   & 2" & PRY\_footpoints\_HI   & N \\
2008-Jan-12 & 14:41:34 & 180$\times$512 & 25   & 2" & PRY\_footpoints\_HI   & N \\
2008-Jan-12 & 15:22:04 & 180$\times$512 & 25   & 2" & PRY\_footpoints\_HI   & N \\
2008-Jan-14 & 12:00:33 & 180$\times$512 & 25   & 2" & PRY\_footpoints\_HI   & N \\
2008-Jan-14 & 13:21:30 & 180$\times$512 & 25   & 2" & PRY\_footpoints\_HI   & N \\
2008-Jan-14 & 14:02:00 & 180$\times$512 & 25   & 2" & PRY\_footpoints\_HI   & N \\
2008-Jan-14 & 14:42:30 & 180$\times$512 & 25   & 2" & PRY\_footpoints\_HI   & N \\
2008-Jan-14 & 21:30:02 & 180$\times$512 & 25   & 2" & PRY\_footpoints\_HI   & N \\
2008-Jan-14 & 22:10:31 & 180$\times$512 & 25   & 2" & PRY\_footpoints\_HI   & N \\
2008-Jan-15 & 00:18:32 & 180$\times$512 & 25   & 2" & PRY\_footpoints\_HI   & N \\
2008-Jan-15 & 11:24:25 & 180$\times$512 & 25   & 2" & PRY\_footpoints\_HI   & N \\
2008-Jan-15 & 12:04:54 & 180$\times$512 & 25   & 2" & PRY\_footpoints\_HI   & N \\
2008-Jan-15 & 12:45:24 & 180$\times$512 & 25   & 2" & PRY\_footpoints\_HI   & N \\
 & & & & \\
\multicolumn{7}{c}{LW detector degradation effects} \\
 & & & & \\
2014-Mar-16 & 12:15:26 & 120$\times$160 & 60   & 2" & ATLAS\_60             & Y \\
2016-Oct-25 & 11:33:15 & 120$\times$160 & 30   & 2" & ATLAS\_30             & Y \\
2018-Dec-28 & 19:10:40 & 120$\times$160 & 60   & 2" & ATLAS\_60             & Y \\
 & & & & \\
\hline
\end{tabular}
\end{center}
\caption{\label{datasets}\small Hinode/EIS observations used in the present work.}
\end{table}

Very careful considerations were made in the choice of the intensity calibration. 
The GDZ calibration was determined monitoring a large number of line intensity ratios 
from the beginning of the EIS mission, and made extensive comparison with high accuracy 
line intensity measurements available in the literature. The HPW calibration attempted 
to improve on the GDZ calibration by relying on extensive plasma diagnostic measurements 
in near-isothermal quiet regions rather than on individual line intensity ratios, by 
tying the EIS calibration to independent measurements from the EUV Variability Experiment 
(EVE -- Woods \etal 2012) and Atmospheric Imaging Assembly (AIA -- Lemen \etal 2012) on
board the Solar Dynamic Observatory (SDO -- Pesnell \etal 2012), and also trying to 
calibrate the SW and LW detectors by making measurements of the \ion[Fe xxiv] 192/255 
line intensity ratio during flares agree with their predicted values. 

No definitive conclusion can be reached regarding which of the two calibrations is most 
accurate. However, the GDZ calibration assumes that the LW channel did not decrease its 
sensitivity with time after 2012, while the HPW calibration assumes a continuous decrease.
A monitoring of the \ion[Fe xiv] 211.32/274.20 line intensity ratio from 2007 to 2020 
indeed shows an initial decrease in relative sensitivity of the LW channel until 2012, 
and a subsequent flattening of the ratio (H.P. Warren, private communications), indicating 
that the GDZ calibration is more accurate after 2012. In this work, we preferred the GDZ 
calibration over the HPW one for three reasons: 1) no other systematic calibration study 
has been published after 2014 that provides a definitive answer on the EIS calibration; 
2) the GDZ calibration is likely more accurate after 2012, and 3) for the sake of 
consistency. We will discuss the effects of the calibration choice in 
Section~\ref{uncertainties}.

Line intensities were determined by summing the number of counts under the line profile
and subtracting the background, determining the latter from regions of the spectrum very 
close to each line devoid of any other line. While such a method provides reasonably good 
estimates of line intensities for isolated lines (and for the 188 doublet), it needs to 
be tested for the 257.26~\AA\ line. In fact, this line is surrounded by several other 
spectral lines which, though resolved, partially mask the wings of the \ion[Fe x] 
transition. While the 257.26~\AA\ line is stronger than all of them, extensive tests 
were run to ensure that the intensities measured in this way are within a few percent 
of intensities calculated by fitting Gaussian profiles to both the \ion[Fe x] and the 
other close-by lines.

\section{Results}
\label{results}

Since this technique simply consists in determining the excess emission in the 257.26~\AA, 
in principle it can be applied to all the EIS observing sequences that include this line, 
a reference \ion[Fe x] line, and a suitable density diagnostic line pair to measure the 
electron density with. While not all EIS observing sequences have all these required 
lines, observations that include them span the entire EIS mission. As such, applying 
this technique to this wealth of data is beyond the scope of this work; here we focus 
on presenting a few examples of this application, and discuss their uncertainties.

\subsection{Magnetic field morphology}

\begin{figure}[!t]
\includegraphics[width=7.0cm,height=8.0cm,angle=90]{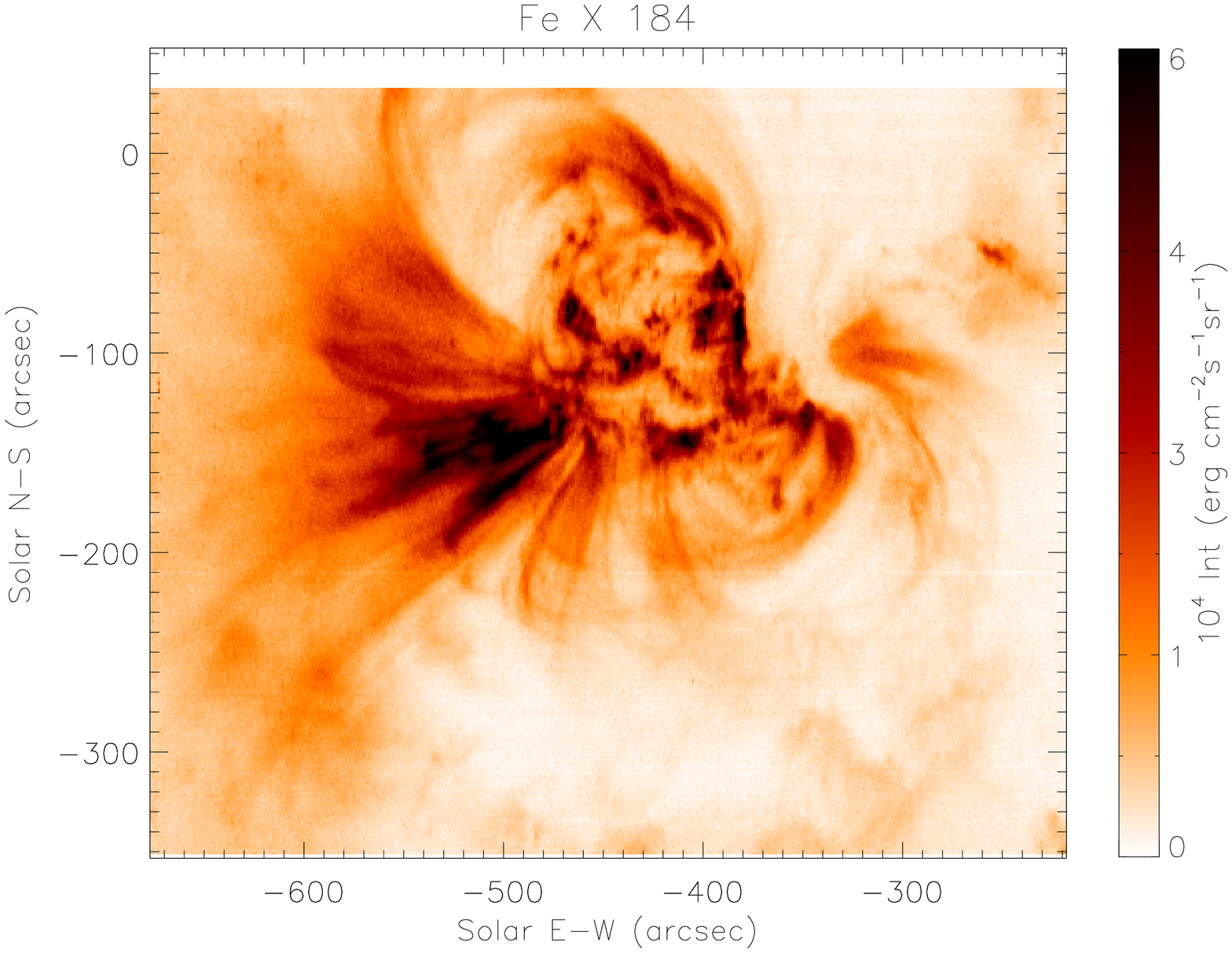}
\includegraphics[width=7.0cm,height=8.0cm,angle=90]{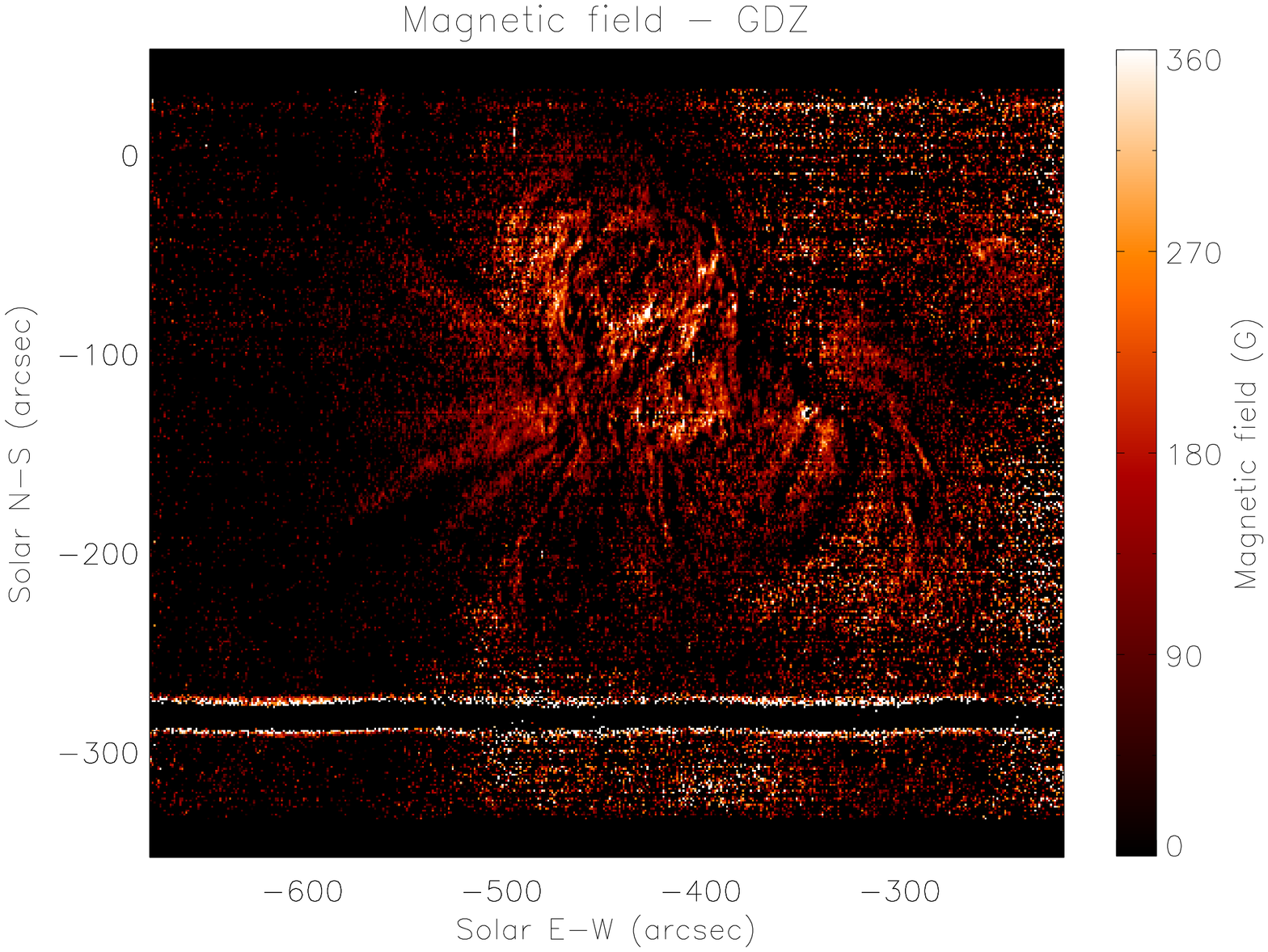}
\includegraphics[width=7.0cm,height=8.0cm,angle=90]{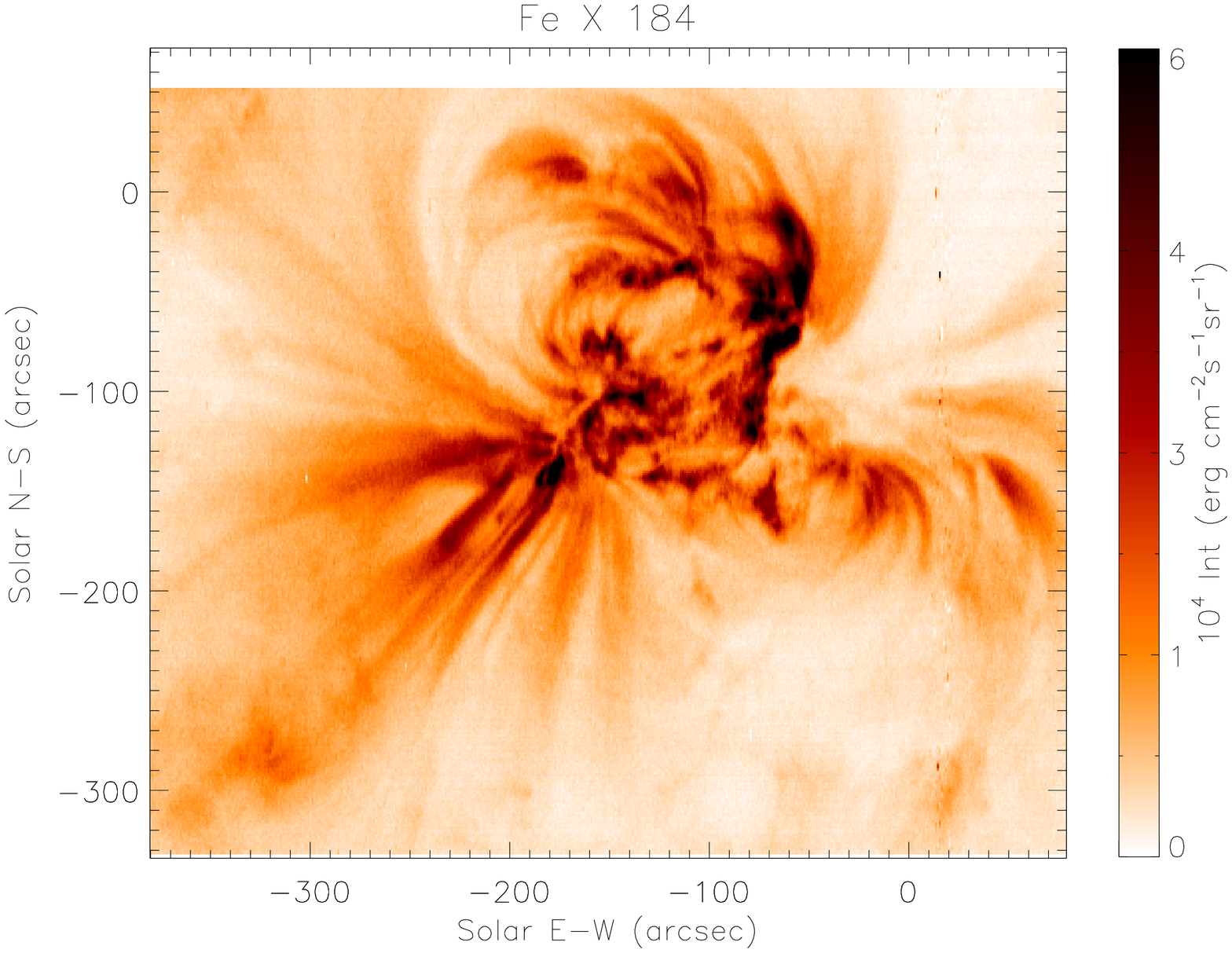}
\includegraphics[width=7.0cm,height=8.0cm,angle=90]{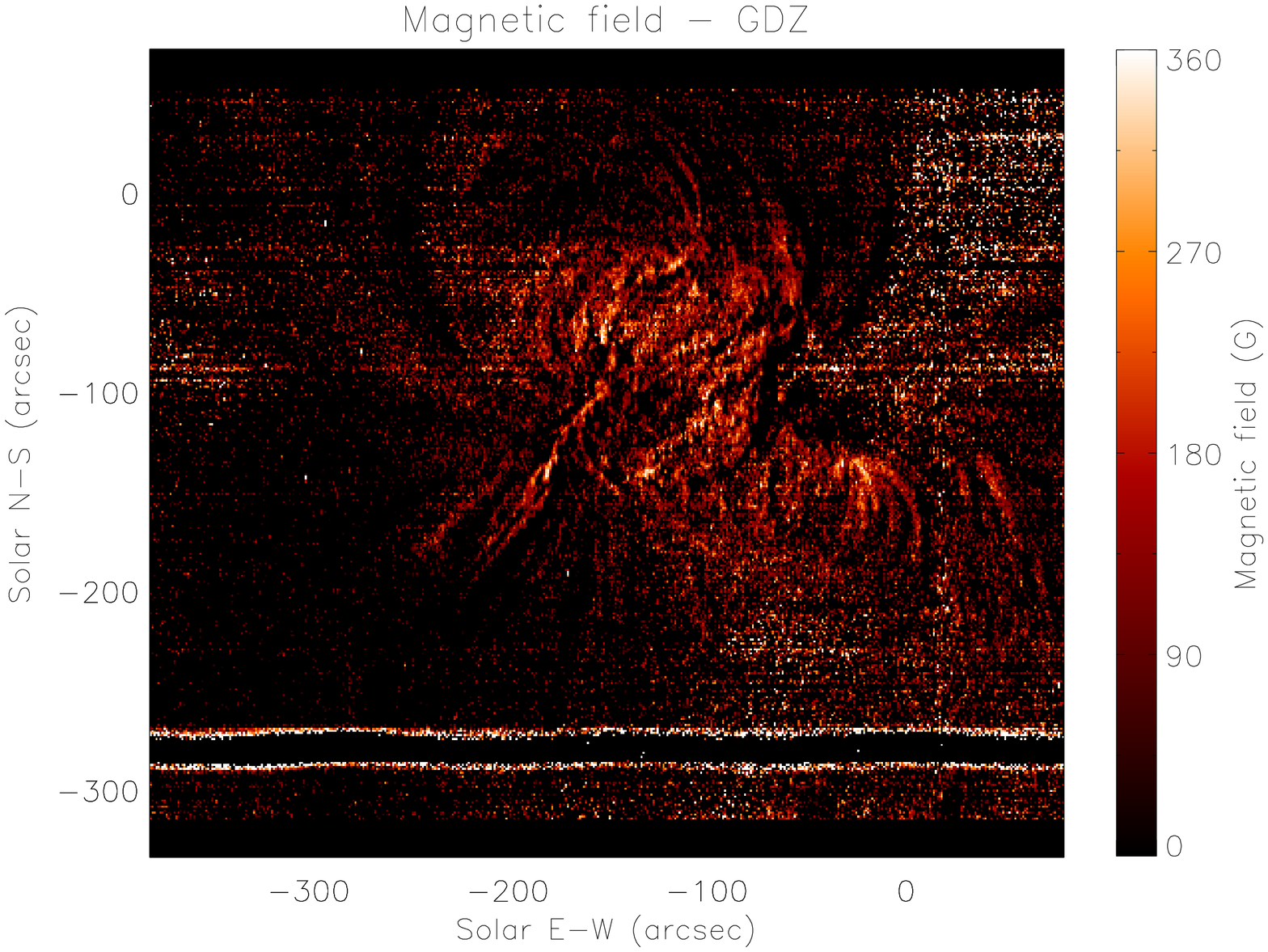}
\includegraphics[width=7.0cm,height=8.0cm,angle=90]{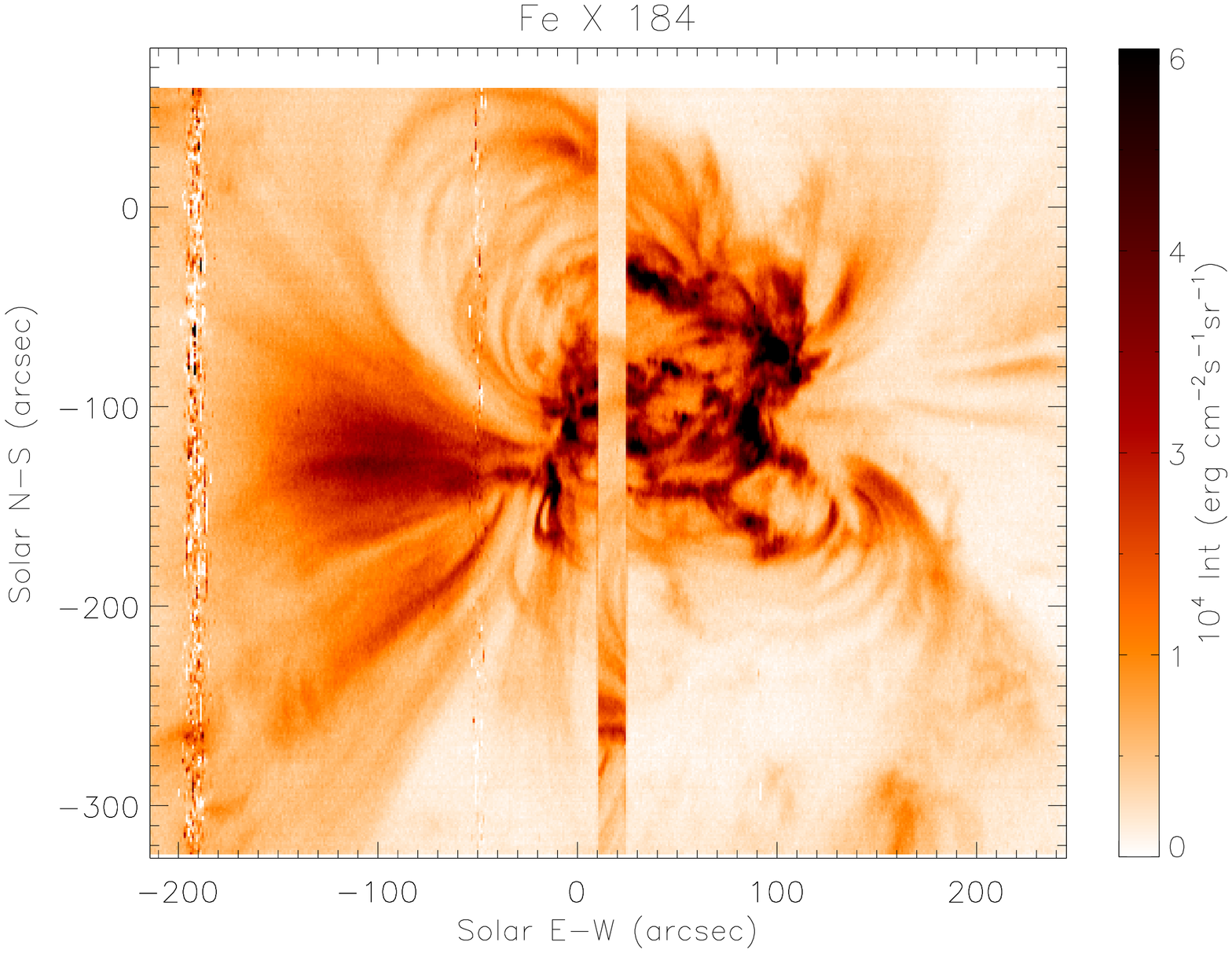}
\includegraphics[width=7.0cm,height=8.0cm,angle=90]{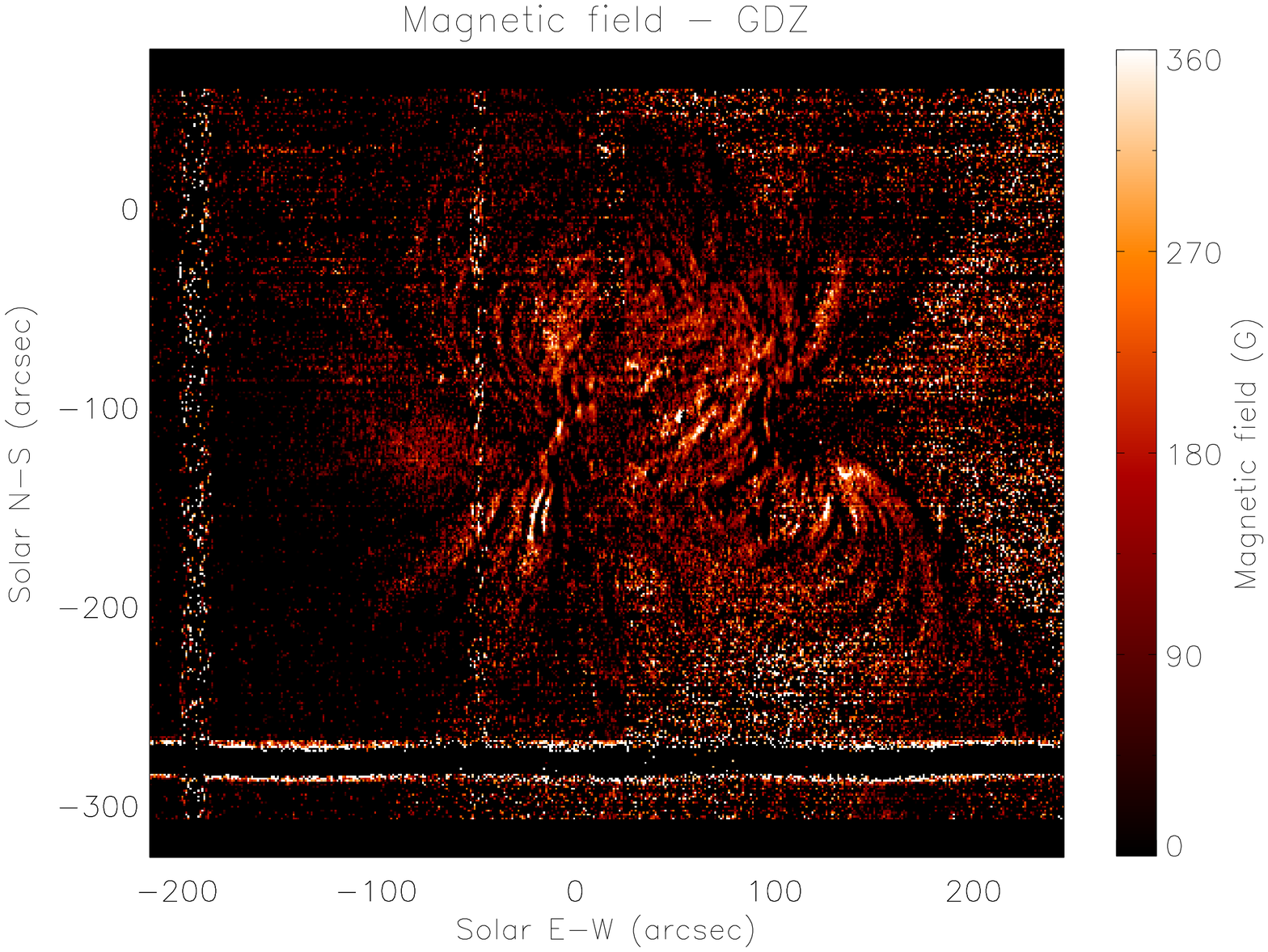}
\caption{AR10978 maps on December 2007: \ion[Fe x] 184.54~\AA\ intensity (left) and magnetic field
strenth (right). Observation days are 10 December (00:19:27~UT, top), 11 December (10:25:42~UT, middle) 
and 12 December (03:26:43~UT, bottom).}
\label{ar1}
\end{figure}

\begin{figure}[!t]
\includegraphics[width=7.0cm,height=8.0cm,angle=90]{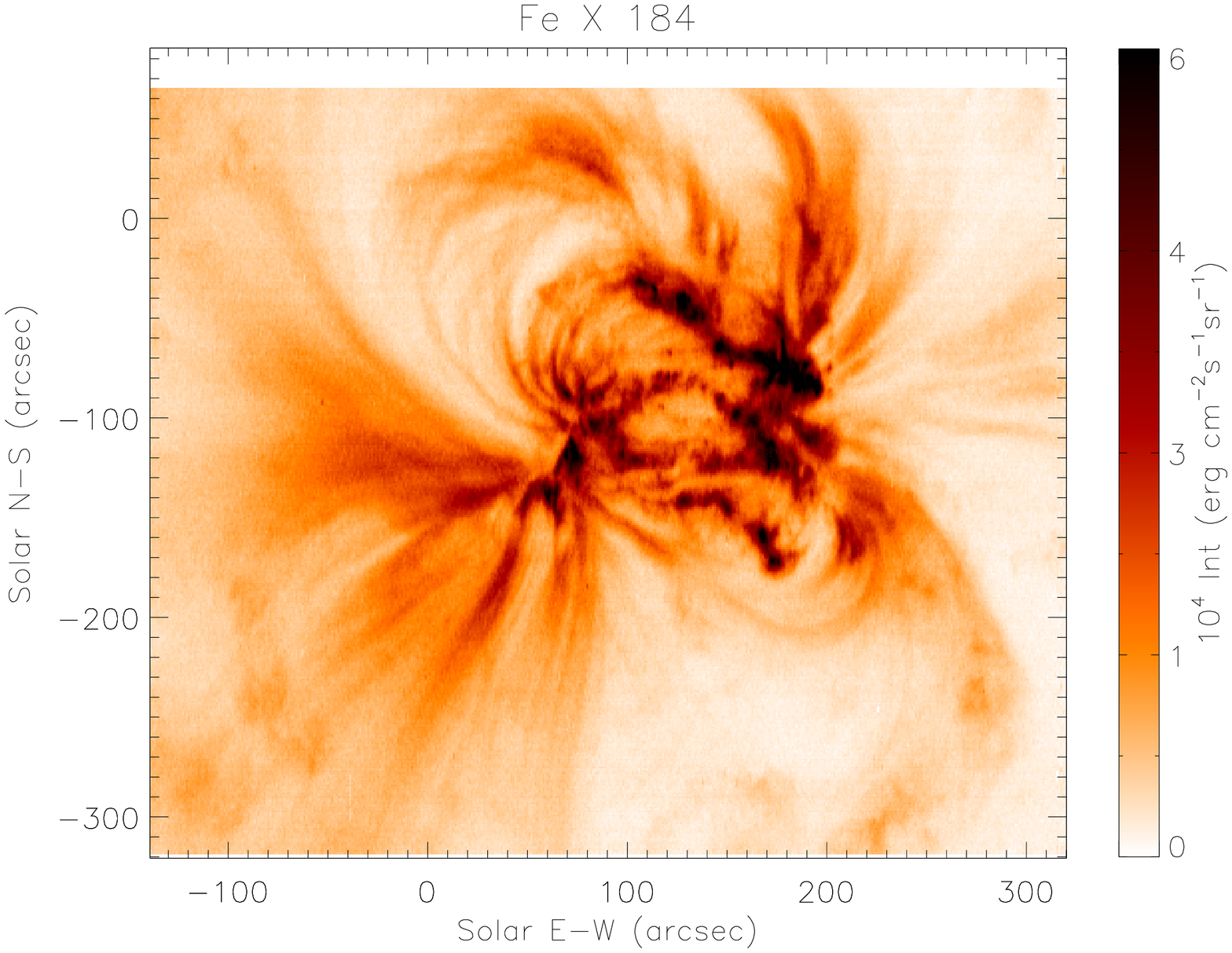}
\includegraphics[width=7.0cm,height=8.0cm,angle=90]{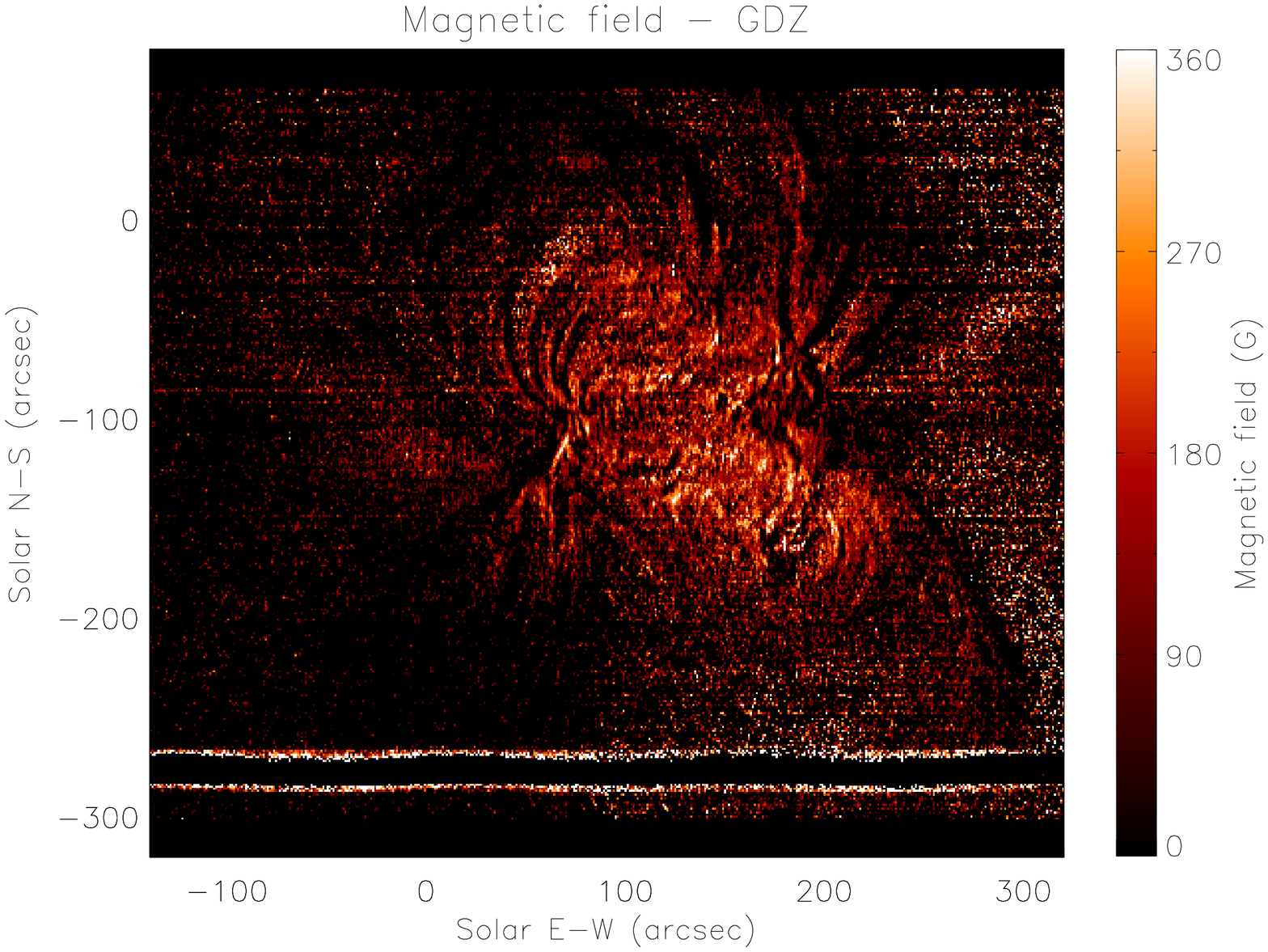}
\includegraphics[width=7.0cm,height=8.0cm,angle=90]{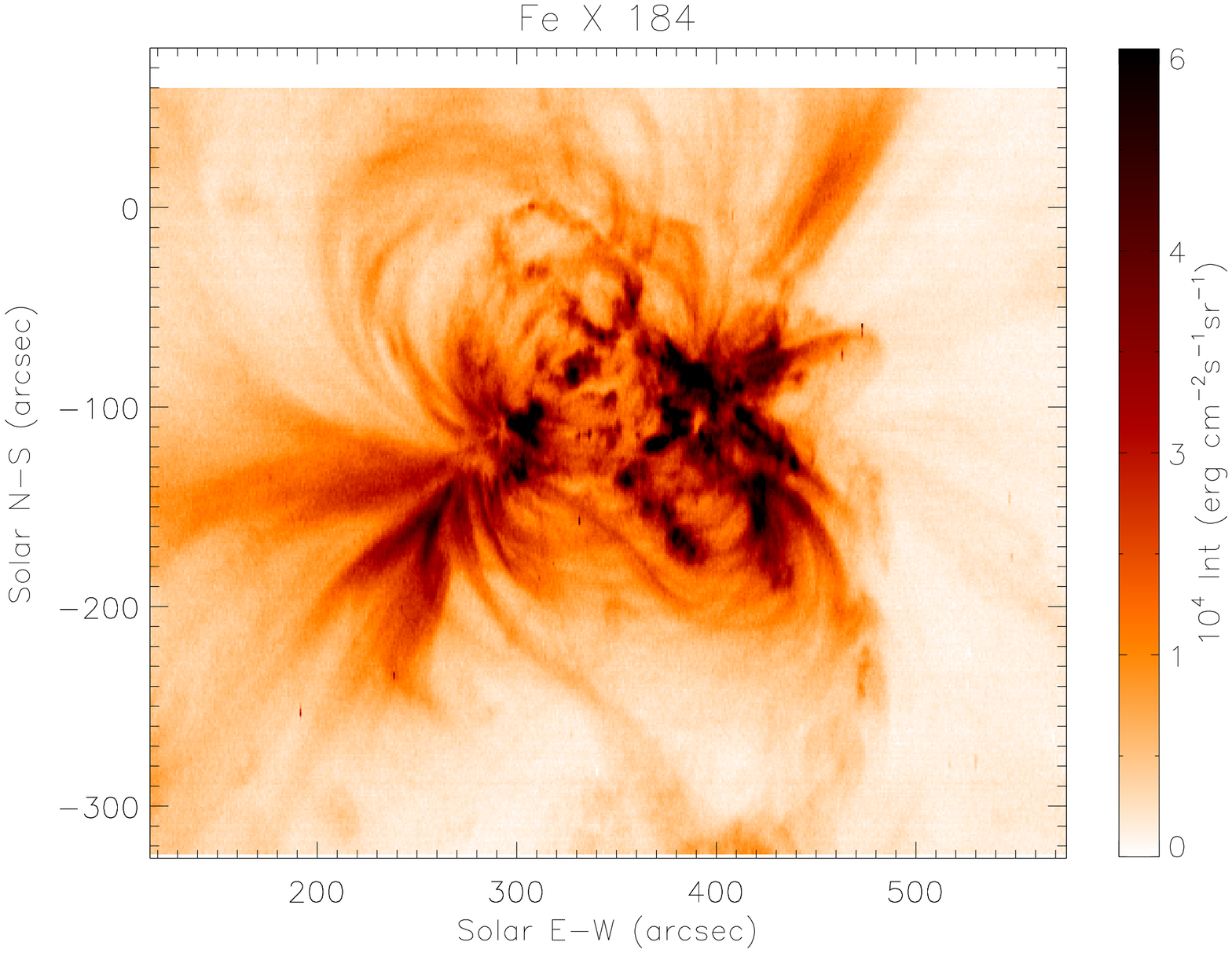}
\includegraphics[width=7.0cm,height=8.0cm,angle=90]{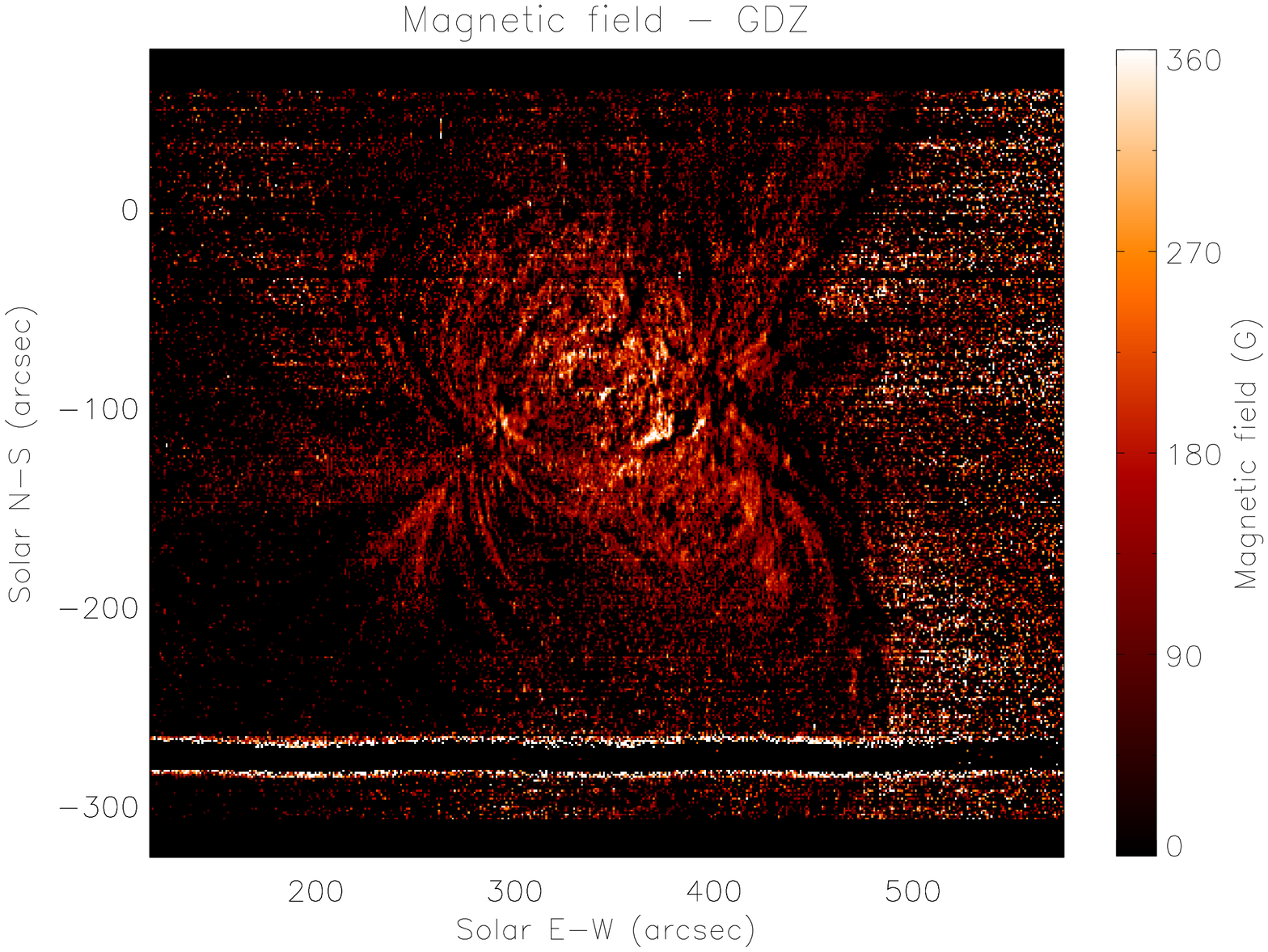}
\includegraphics[width=7.0cm,height=8.0cm,angle=90]{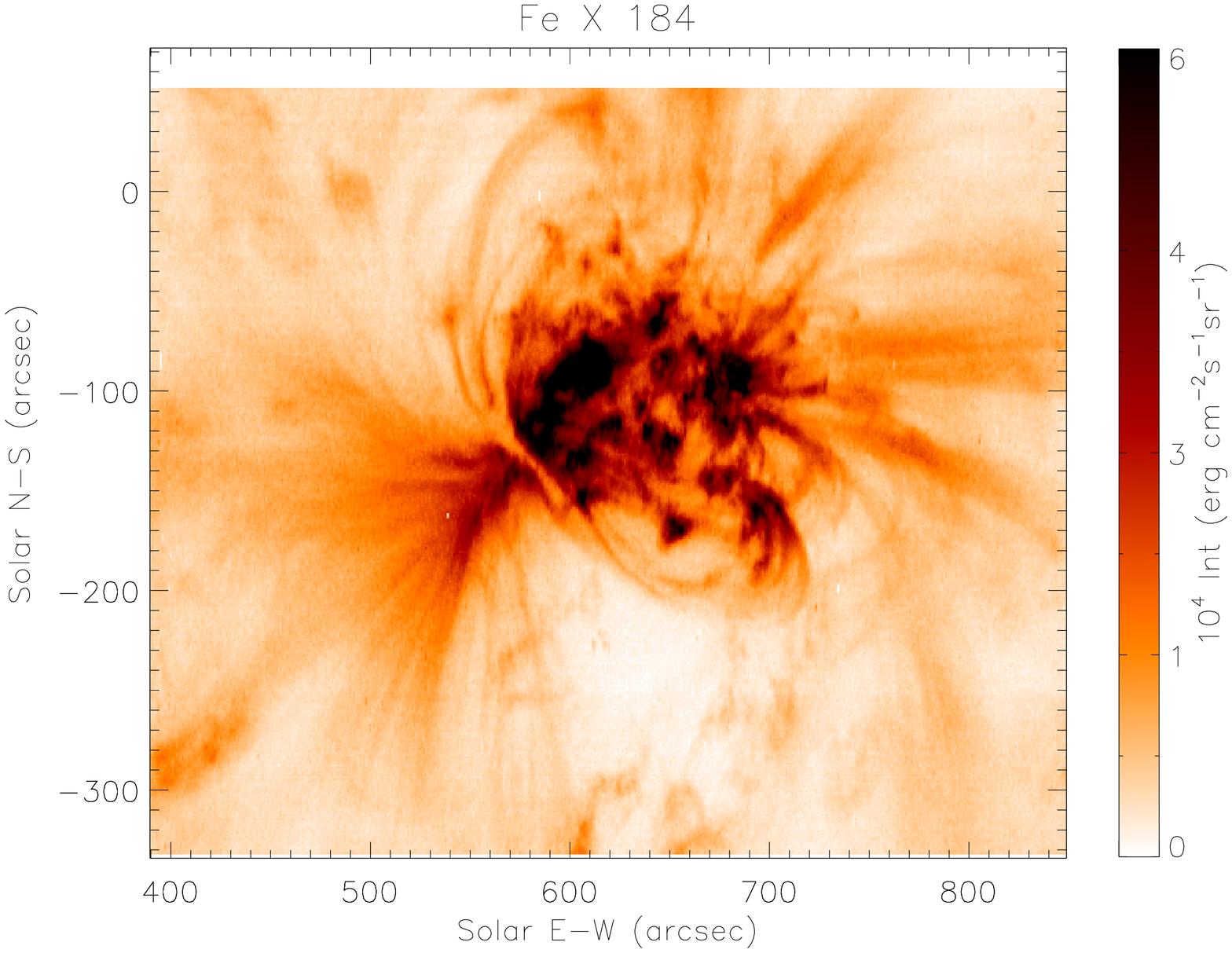}
\includegraphics[width=7.0cm,height=8.0cm,angle=90]{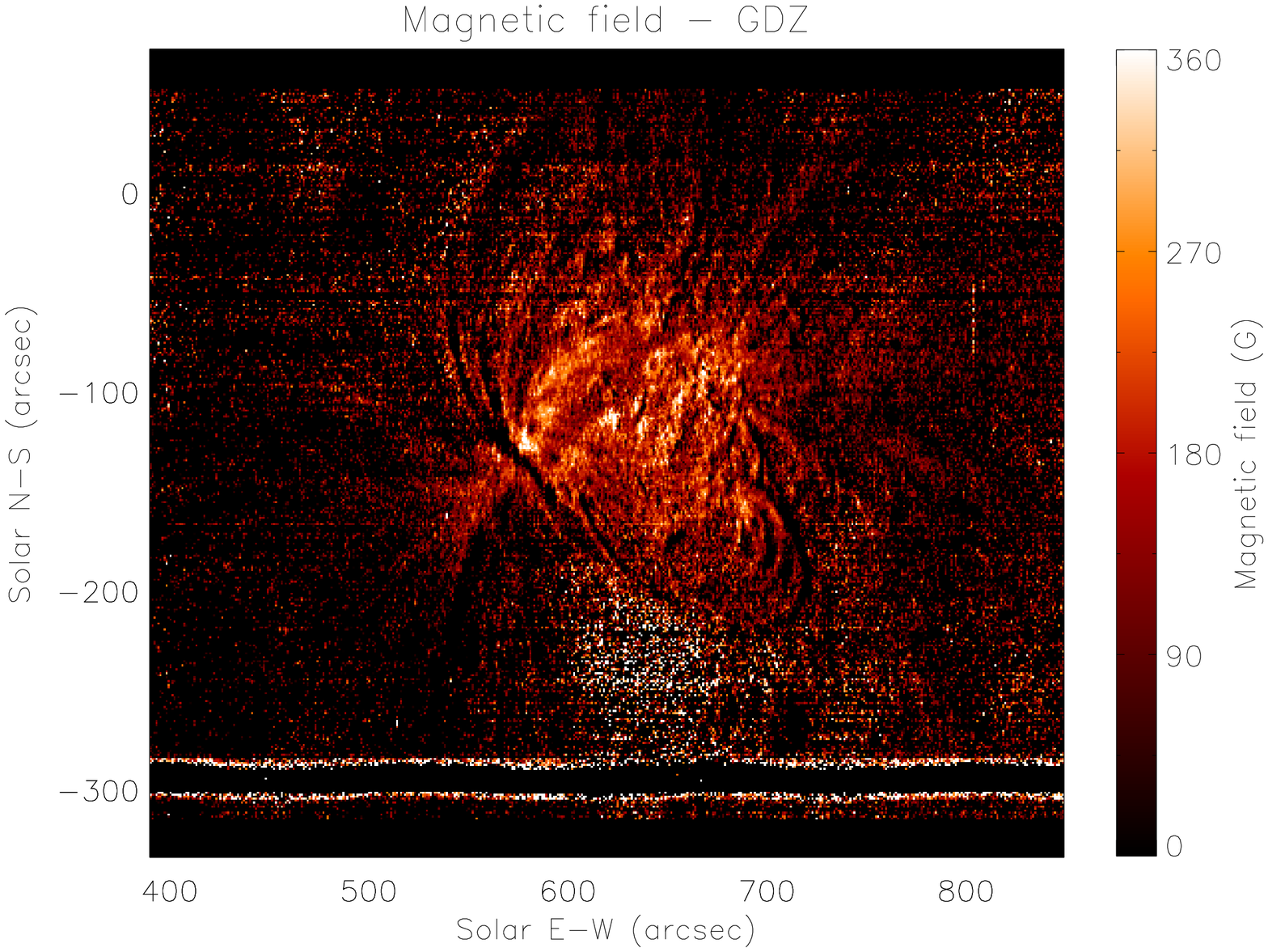}
\caption{AR10978 maps on December 2007: \ion[Fe x] 184.54~\AA\ intensity (left) and magnetic field
strenth (right). Observation days are 12 December (11:43:36~UT, top), 13 December (12:18:42~UT, middle) 
and 15 December (00:13:49~UT, bottom).}
\label{ar2}
\end{figure}

\begin{figure}[!t]
\includegraphics[width=7.0cm,height=8.0cm,angle=90]{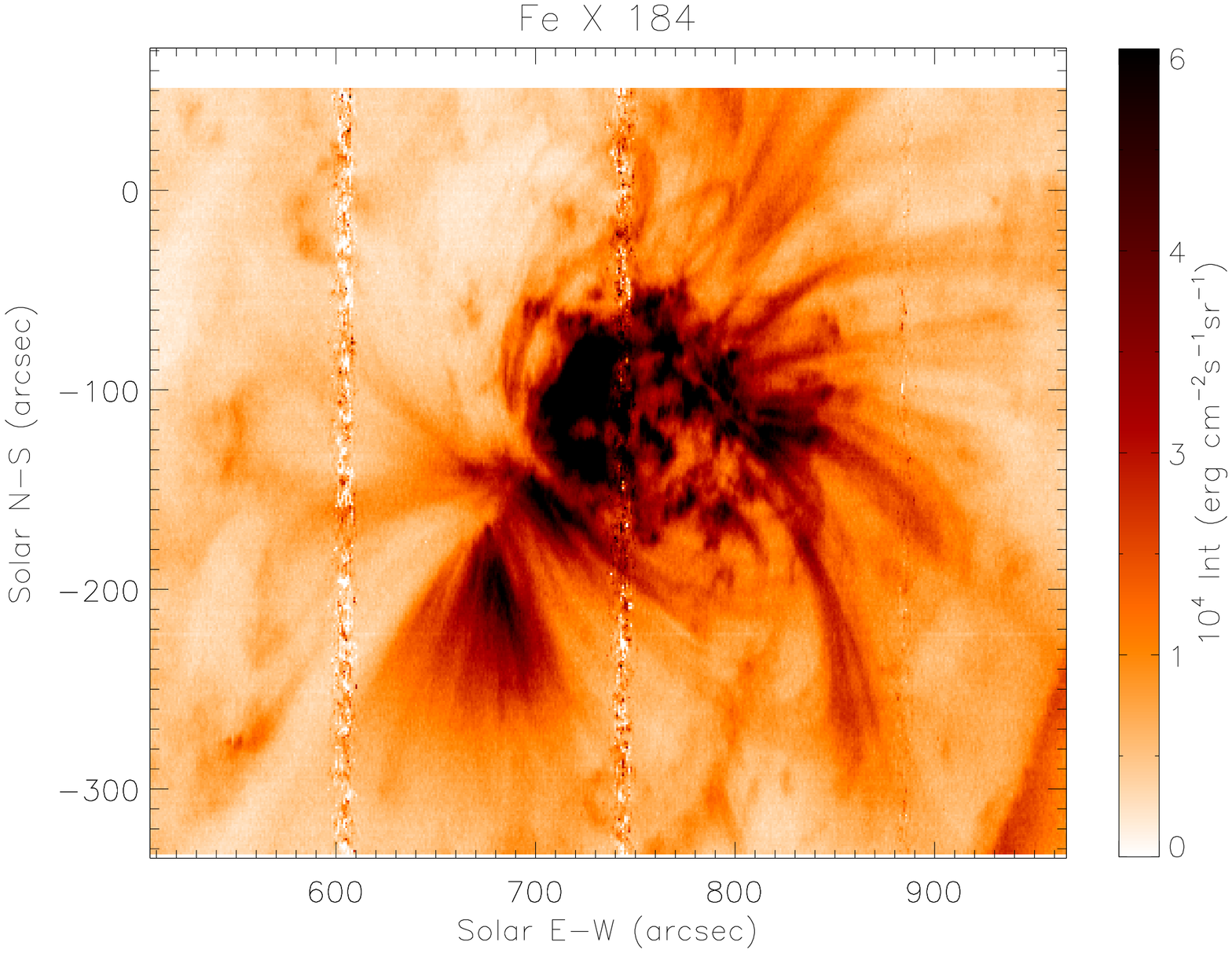}
\includegraphics[width=7.0cm,height=8.0cm,angle=90]{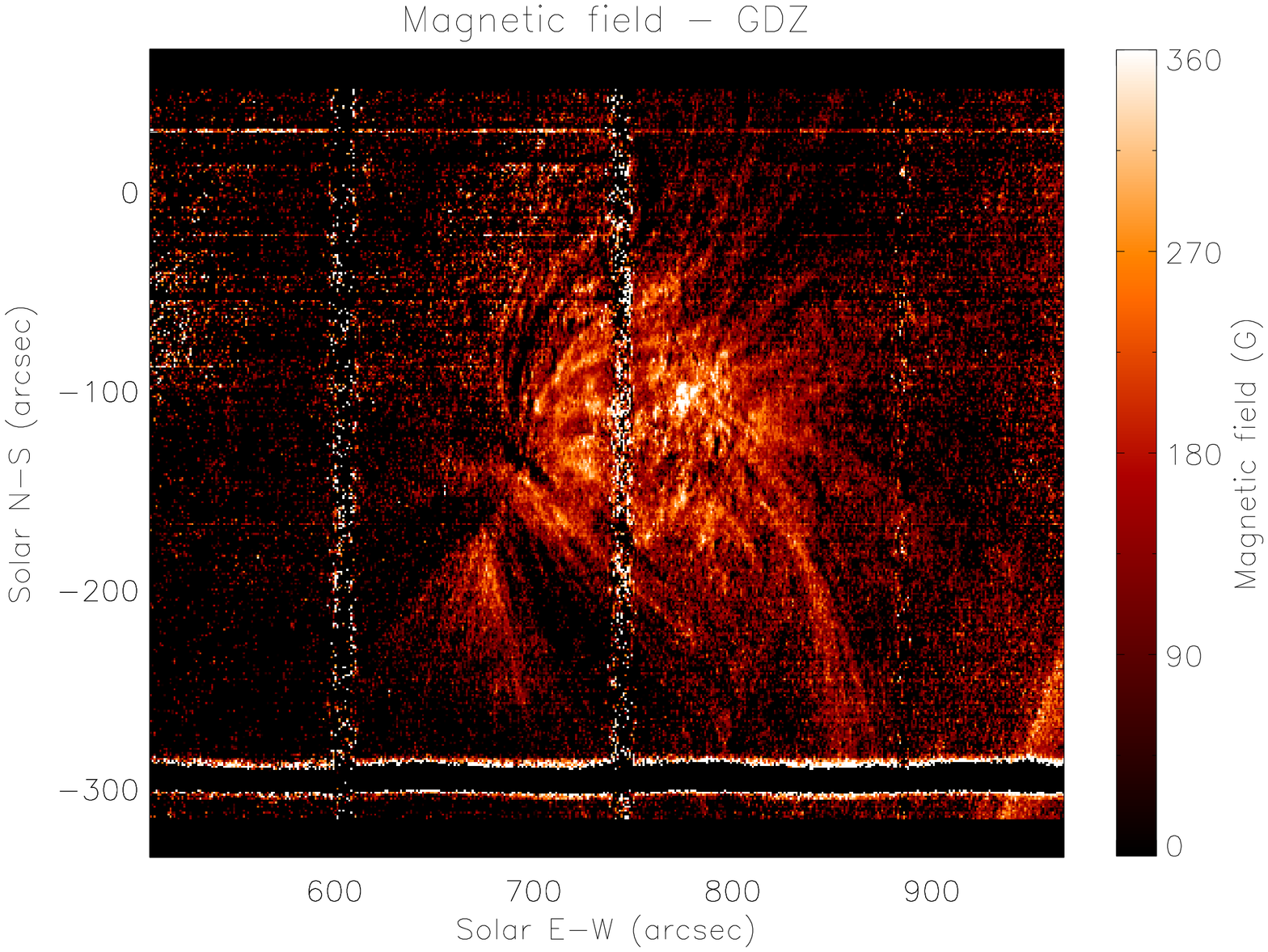}
\includegraphics[width=7.0cm,height=8.0cm,angle=90]{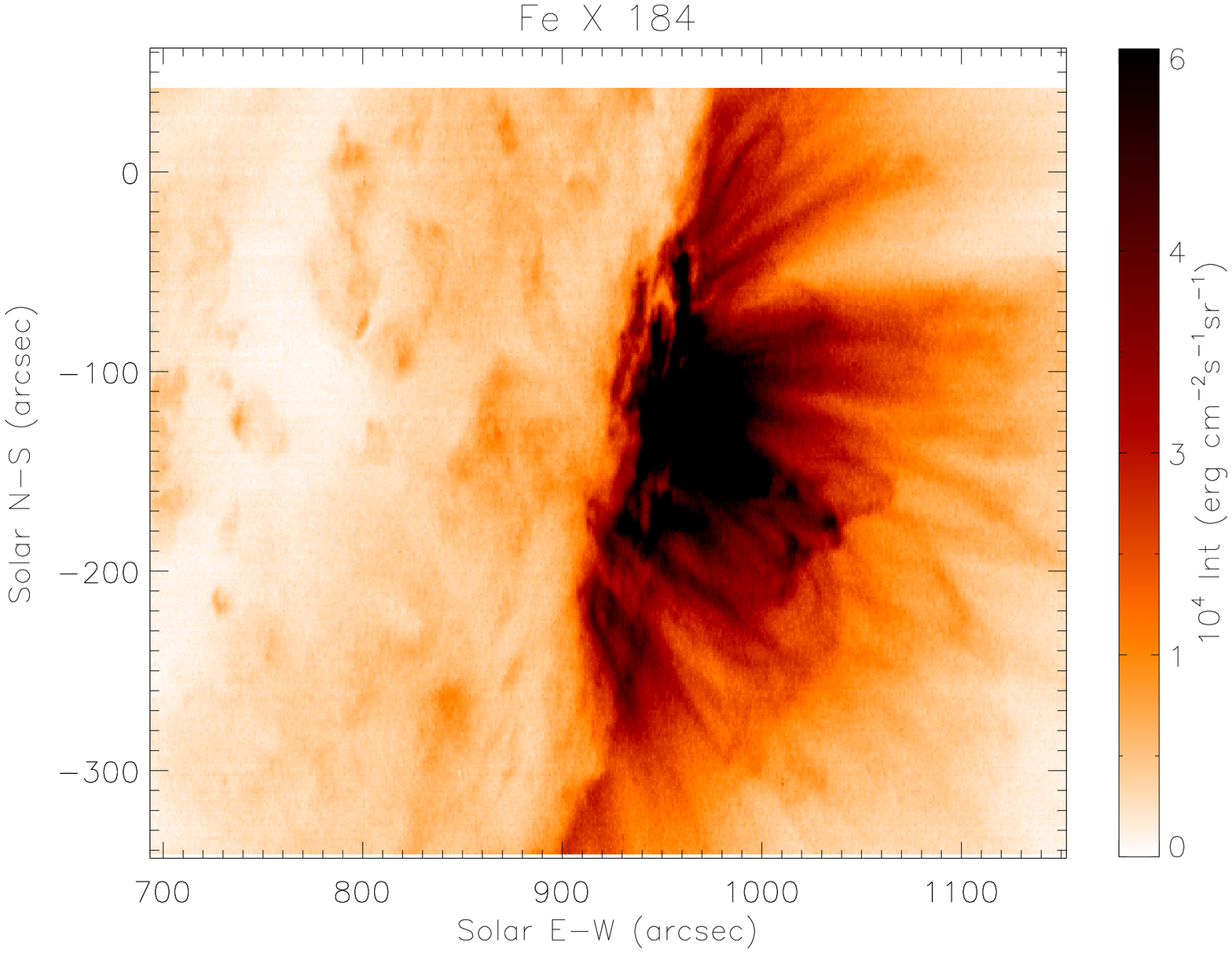}
\includegraphics[width=7.0cm,height=8.0cm,angle=90]{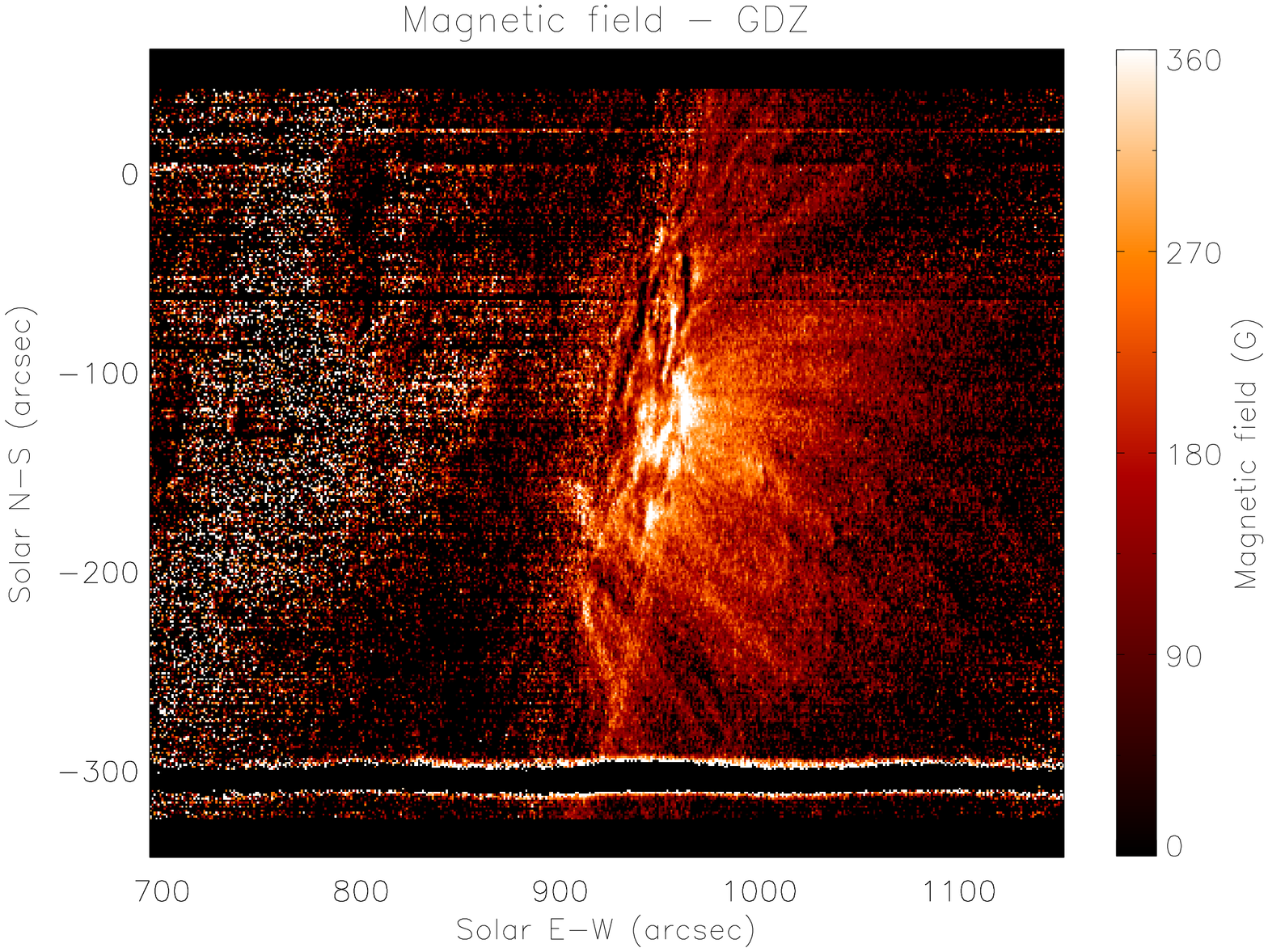}
\includegraphics[width=7.0cm,height=8.0cm,angle=90]{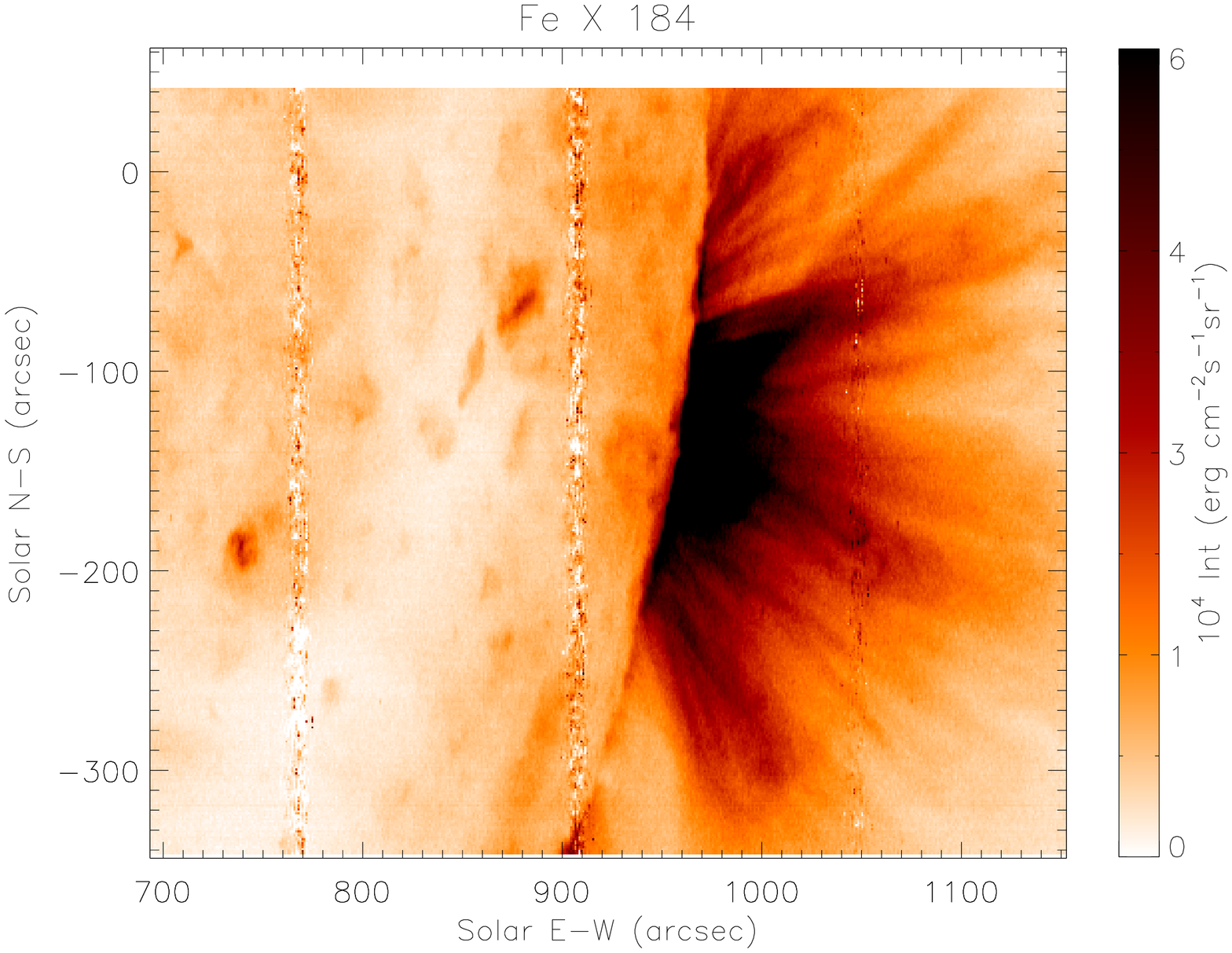}
\includegraphics[width=7.0cm,height=8.0cm,angle=90]{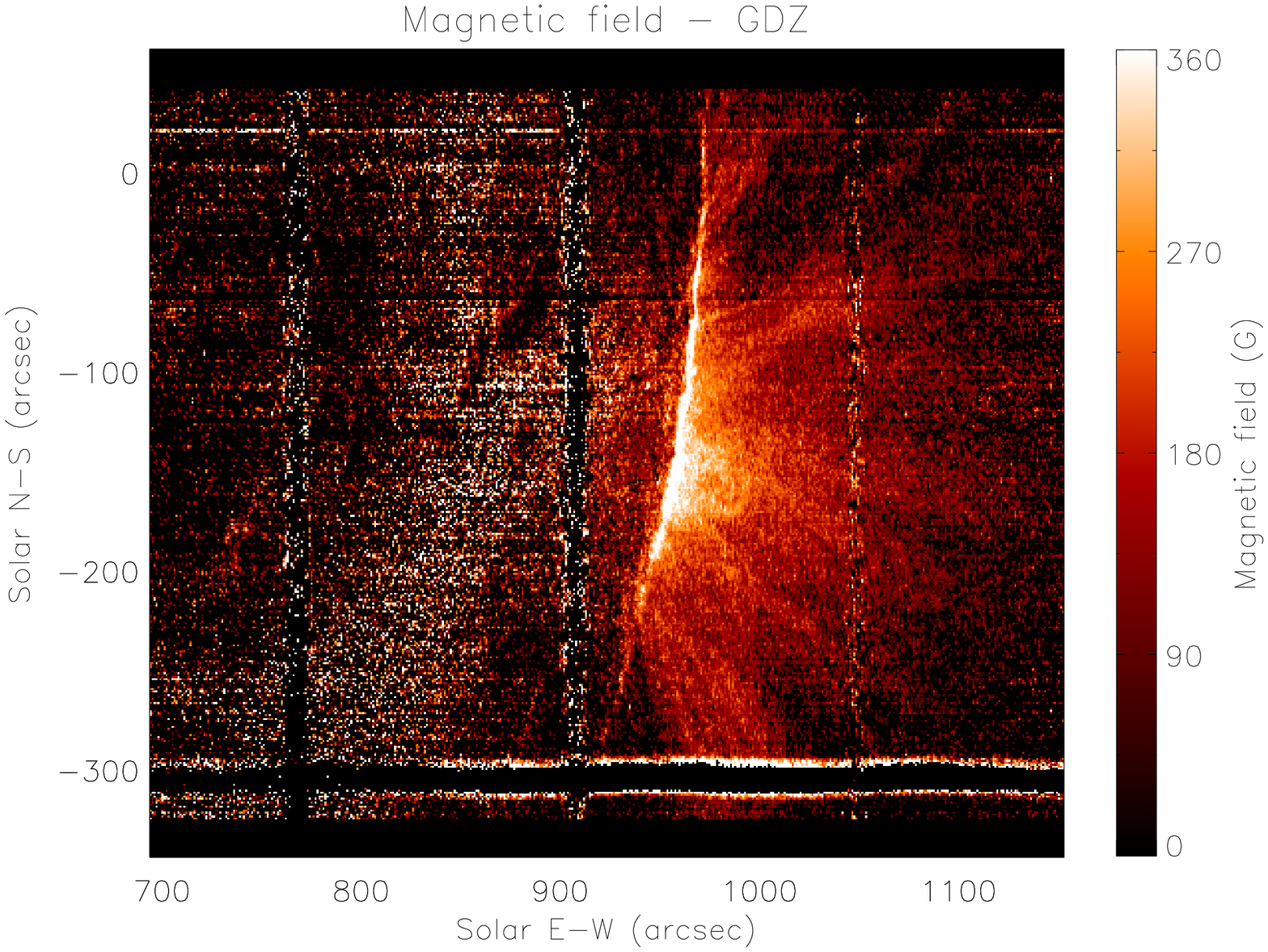}
\caption{AR10978 maps on December 2007: \ion[Fe x] 184.54~\AA\ intensity (left) and magnetic field
strenth (right). Observation days are 15 December (18:15:49~UT, top), and 18 December (00:10:49~UT, 
middle and 18:13:41~UT, bottom).}
\label{ar3}
\end{figure}

An example of the application of the \ion[Fe x] magnetic field diagnostic technique is 
shown in Figures~\ref{ar1} to \ref{ar3}. This set of observations, carried out between 
2007 December~10 and 18, follows AR10978 as it rotated on the solar disk until reaching 
the west limb. Despite the size and complexity, this active region was relatively quiet 
and hosted only a few C-class flares, although none of them took place while the 
observations were taken. Figures~\ref{ar1} to \ref{ar3} show a snapshot of the \ion[Fe x] 
184.54~\AA\ intensity map on the left column and the magnetic field map obtained with the
present MIT diagnostic technique on the right 
column for each of the nine observation times. There are a few things to notice.

First, the magnetic field structures shown in the figures closely follow the intensity 
maps of the \ion[Fe x] line, as expected both on the grounds of better SNR, and because 
the magnetic field confines the active region plasma maintaining it at higher density 
and temperature than the surrounding ambient plasma. Observations closer to disk center 
allow the reconstruction of several different loop structures connecting magnetic field of 
opposite polarities, although the tallest loops become too faint as they rise to large
altitude, such as those in the SW portion of AR10978 in Figure~\ref{ar1}, so that they
can not be observed for the entirety of their length.

Second, even if the observations were taken only a few hours apart from each other, the 
magnetic field shows significant evolution, both in morphology and in strength, with 
magnetic loops changing position, appearing or disappearing from one observation to the 
next. This is consistent with the variability of line intensities, and directly links the 
evolution of the field strengths with the plasma properties inside magnetic structures.

Third, the strength of the magnetic field ranges from $\approx$100~G to more than 300~G, 
being stronger in low-lying loops and weaker in larger loops. However, it is unclear 
whether the lower strength in large loops (leading to disappearance of the loop 
themselves at large heights) is due to a real weakening of the field with height, 
or to a lower SNR. During the transit of AR10978 from disk center to the limb, the 
magnetic field strength steadily increases, consistently with the active region 
beginning to be more active and starting to host small C-class flares from December~13 
(three C-class flares, the strongest of which a C4.5 flare) until it turned behind the 
limb on December~19.

The variability of the magnetic field strength with height can be best monitored in 
Figure~\ref{ar3}, where AR10978 is observed close to, or at the limb. In these images, 
the strongest magnetic field (reaching and exceeding 300~G) is concentrated in the 
lowest-lying structures, closest to the surface, while taller loops have weaker fields, 
in the 100-250~G range. The tallest loops visible in the \ion[Fe x] 184.54~\AA\ image 
are barely visible in the magnetic field strength image, indicating that their magnetic 
field is either too weak to be detected, or the SNR is too low to allow this diagnostic 
technique to be effective. However, the very few closed loop structures whose magnetic 
field strength is measurable at the limb for seemingly the whole loop length show a 
slow varying magnetic field, which becomes weaker with height.

\subsection{Magnetic field evolution}

\begin{figure}[!t]
\includegraphics[width=10.0cm,height=16.0cm,angle=90]{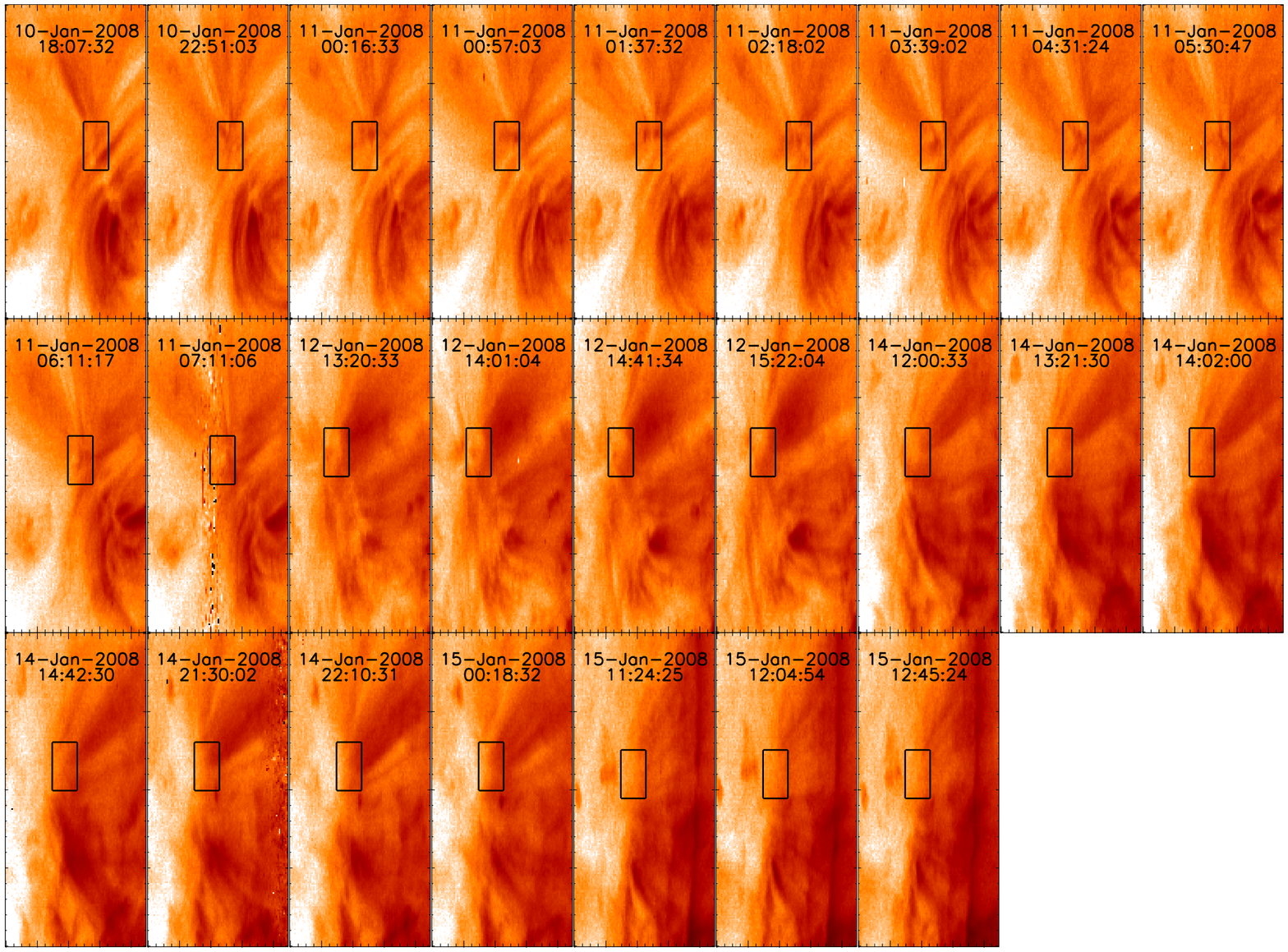}
\includegraphics[width=9.0cm,height=16.0cm,angle=90]{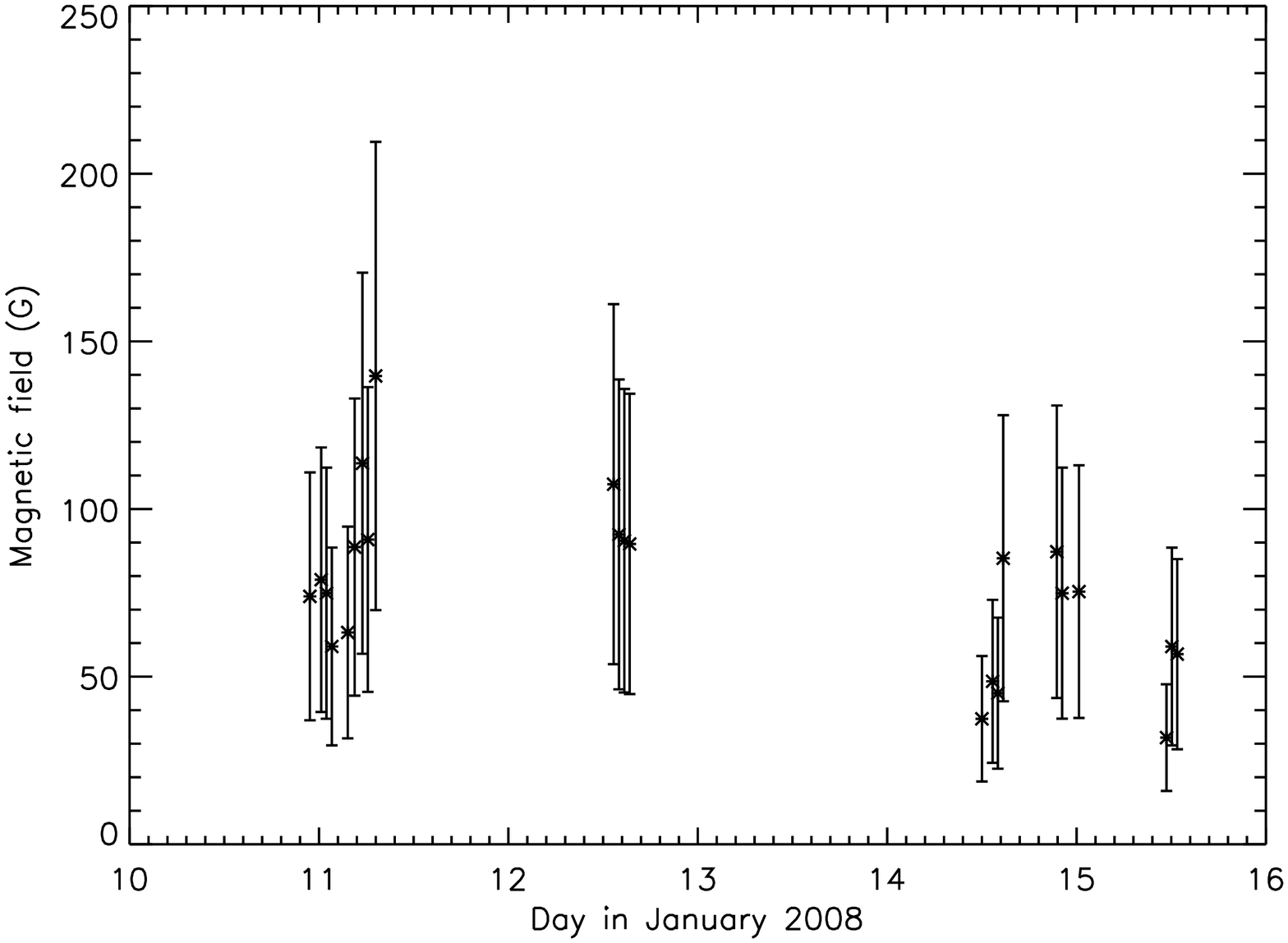}
\caption{Magnetic field measurement for a weak active region observed between January~10
to 15, 2008. Intensities have been averaged over the reported boxes for each of the 
observations, and the magnetic field measurement has been carried out on the 
averaged intensities. {\bf Top panel:} \ion[Fe x] 184.54~\AA\ intensity maps of a 
portion of the active region including the footpoints of fanning loops. {\bf Bottom
panel:} magnetic field strength measurements as a function of time.}
\label{evol1}
\end{figure}

Magnetic field strength maps also allow us to monitor the short term evolution of individual
structures in an active region. An example of this is reported in Figure~\ref{evol1}, where
a weak active region with no number, which was trailing AR10980 at the solar equator, was
repeatedly observed from January 10 to 15, 2008. During this time, the active region was 
observed for a total of 25 times, and showed a marked evolution which led to no flares.

The high resolution of the EIS spectrometer allows us to identify individual plasma 
structures and monitor their evolution with time: we have identified the footpoint of a
system of active region fanning loops, which is highlighted in Figure~\ref{evol1} (top)
by the black rectangle, and measured the magnetic field of the selected box as a function
of time. The magnetic field of this region looked very weak, as a pixel-by-pixel map of 
the coronal magnetic field of this region indicated the presence of a measurable field, 
with a very poor SNR. In order to increase the SNR and attempt to measure the magnetic 
field strength in such a weak region, we have summed all the counts for each spectral 
line within the box and applied the magnetic field diagnostic technique to the total 
line intensities.

Results are shown in Figure~\ref{evol1} (bottom). The uncertainties of each measurement
are given by the uncertainty in the relative calibration between the EIS LW and SW channel,
estimated to be 50\%. The magnetic field is very weak, on the
order of 30-150~G, and shows a vague tendency to decrease with time. Results tend to show 
some variability within the same cluster of observations (a couple of observations gave 
zero magnetic field strength), indicating that we are sampling the sensitivity limits of 
the present magnetic field diagnostic technique. This means that it will be very difficult 
to measure magnetic fields in the quiet Sun and even less in coronal holes, where magnetic 
field strengths are smaller; the only places where such a measurement may be attempted 
with some hope of success are low-latitude locations at the solar limb, where the very 
long line of sight intercepts more plasma, increasing the SNR. 

Another difficulty at carrying out such measurements is the evolution of the plasma
structures themselves, and the different viewing angle as they rotate on the solar disk.
Both properties makes the identification of an individual structure and its monitoring 
over long periods of time difficult and adds to the uncertainty of the measurement.

\section{Uncertainties}
\label{uncertainties}

\subsection{Energy level separation}

Despite the great potential for magnetic field diagnostics, the present technique has several
uncertanties of different nature. The first and most important is the intrinsic uncertainty
in the calculation of the Einstein coefficient $A_{MIT}$ for the magnetically-induced transition.
As discussed by Si \etal (2020a), to first order we can assume

\begin{equation}
A_{MIT} \propto \frac{B^2}{{\left({\Delta E}\right)^2}}
\label{a_mit}
\end{equation}

\noindent 
where $\Delta E$ is the energy separation between the two \tm[4 D 5/2,7/2] levels generating
the 257.26~\AA\ doublet. Measuring this energy separation is difficult because it is very 
small, and there are only a few spectral lines available that can be utilized, all very close 
in wavelength. Separating the two lines at 257.26~\AA\ is essentially impossible as their 
separation is just a few m\AA, way below the resolution of the EIS instrument; also the line
width in the corona is way larger than this seeparation. 

$\Delta E$ can be measured using transitions where the \tm[4 D 5/2,7/2] levels are the
lower levels of transitions coming from the same upper level forming a doublet, as used 
by Judge \etal (2016). A few lines are available for this purpose. No laboratory or solar 
observations are available for a doublet at around 3500~\AA, which would provide a 
separation of 0.65~\AA, the 2935~\AA\ doublet, and another doublet at 1611~\AA. Only a 
few observations in the UV are available for the only other three transitions remaining:
at around 1918~\AA, 1603~\AA, and 1028~\AA; the separation of the two lines being larger
at longer wavelengths. Judge \etal (2016) utilized S082 spectra taken at the solar limb 
to measure the 1603~\AA\ doublet separation through double Gaussian fitting, finding 
$\Delta E =3.7\pm 2.9$~cm$^{-1}$, so that the uncertainty is around 80\%. This uncertainty 
makes the technique capable of giving only the order of magnitude of the magnetic field strength. 

More recently, Landi \etal (2020) have utilized the deep exposure of the quiet solar limb 
made with the high-resolution SoHO/SUMER spectrometer (Wilhelm \etal 1995) that provided 
the SUMER off-disk spectral atlas (Curdt \etal 2004); in this observation, the \ion[Fe x] 
1463.49~\AA\ was used to constrain the line width for all \ion[Fe x] lines, helping reduce 
the $\Delta E$ uncertainty. Landi \etal (2020) determined a value of $2.29\pm 0.50$~cm$^{-1}$. 
This uncertainty, at 20\% level, greatly reduces the intrinsic uncertainty of any measurement 
of the magnetic field (see Equation~\ref{a_mit}).

\subsection{Electron density}

In order to determine the $MIT/M2$ branching ratio from the observations, it is 
necessary to determine the plasma electron density independently. This can in principle
be done using \ion[Fe x] transitions, whose emission is generated by the same plasma
emitting the 257.26~\AA\ line. Si \etal (2020b) suggest the use of the two strong
\ion[Fe x] transitions at 175.26~\AA\ and 174.53~\AA: this choice is the most sensible,
but these two lines are observed at the edge of the EIS SW detector where the sensitivity
is very low, and their intensities are highly uncertain. No other \ion[Fe x] line ratios 
with bright lines with high SRN is available in the EIS wavelength range. The next best 
solution is to utilize lines from ions formed at similar temperatures, and the best 
choice is \ion[Fe xi], as both \ion[Fe x,xi] are ions formed in the corona only, while 
\ion[Fe ix] has a strong contribution from colder, upper transition region plasmas which 
makes its emission contaminated by plasmas likely not contributing to \ion[Fe x]. Besides, 
\ion[Fe ix] density sensitive lines in the EIS spectrum are weaker and more difficult to 
observe and disentangle from nearby transitions. On the contrary, \ion[Fe xi] provides 
strong, isolated lines that are routinely observed by the EIS spectrometer: the 
182.17~\AA\ and the 188.22+188.30~\AA\ doublet. The latter is actually a partially 
resolved doublet, but it is very strong and sufficiently isolated to make it very 
easy to measure. 

\begin{figure}[!t]
\includegraphics[width=9.0cm,height=16.0cm,angle=90]{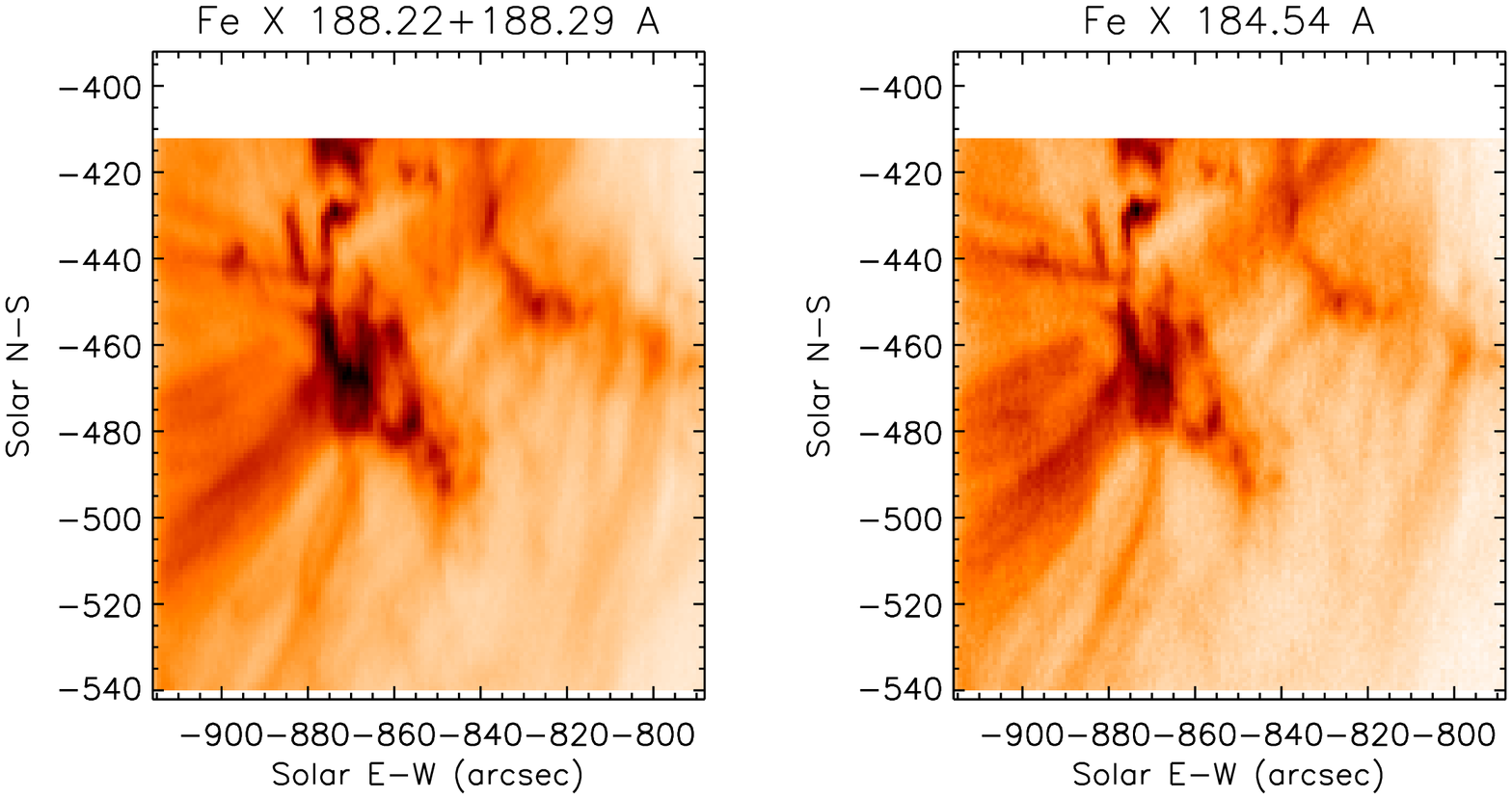}
\includegraphics[width=9.0cm,height=16.0cm,angle=90]{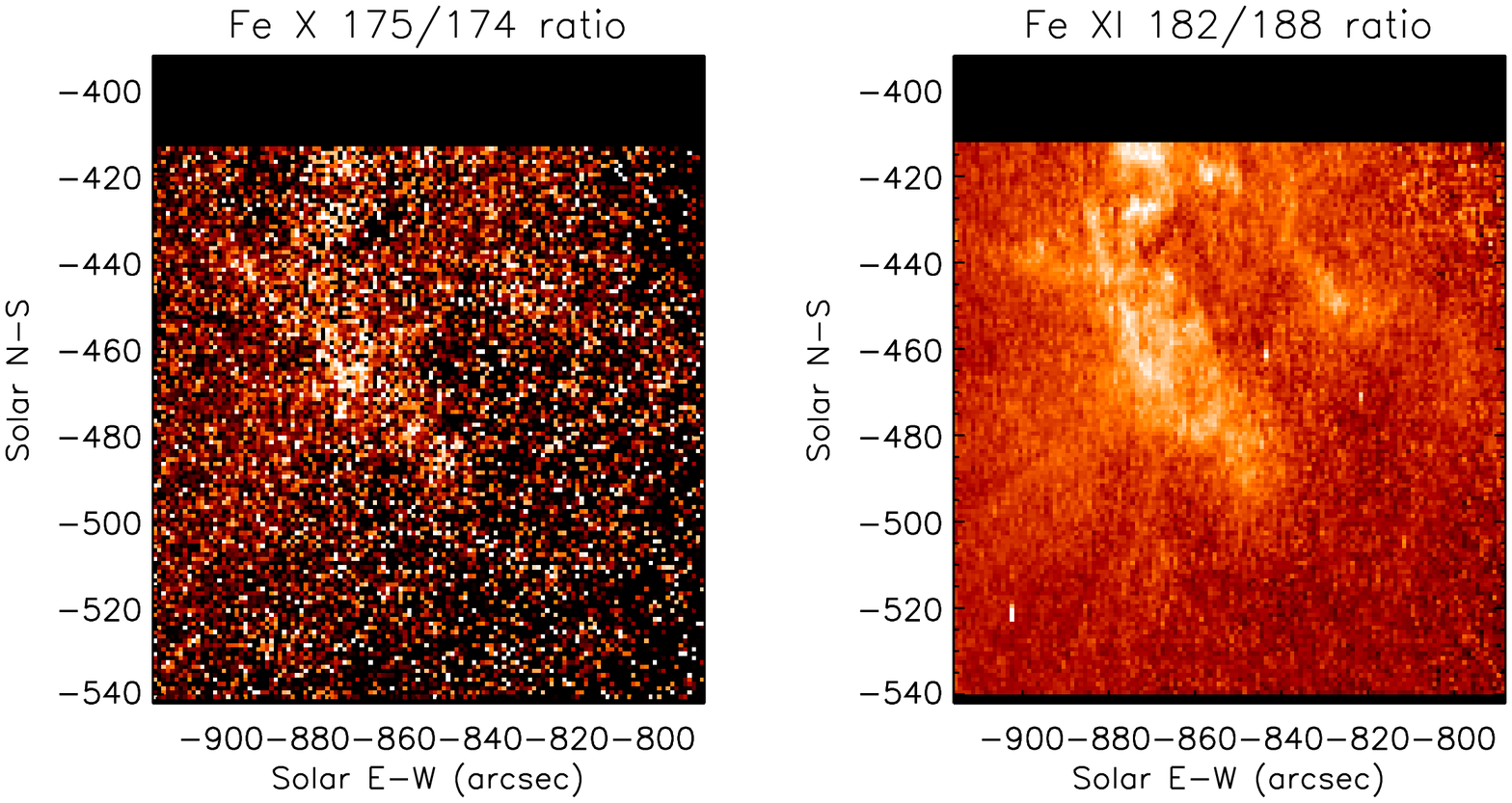}
\caption{EIS observations of AR10960 observed on June 2, 2007. {\bf Top:} intensity
maps obtained with the \ion[Fe xi] 188.2~\AA\ doublet and the \ion[Fe x]
184.54~\AA\ line. {\bf Bottom left:} Density map obtained with the \ion[Fe x]
175.26/174.54 intensity ratio; {\bf Bottom right:} Density map obtained with
the \ion[Fe xi] 182/188 intensity ratio.}
\label{dens1}
\end{figure}

\begin{figure}[!t]
\includegraphics[width=9.0cm,height=16.0cm,angle=90]{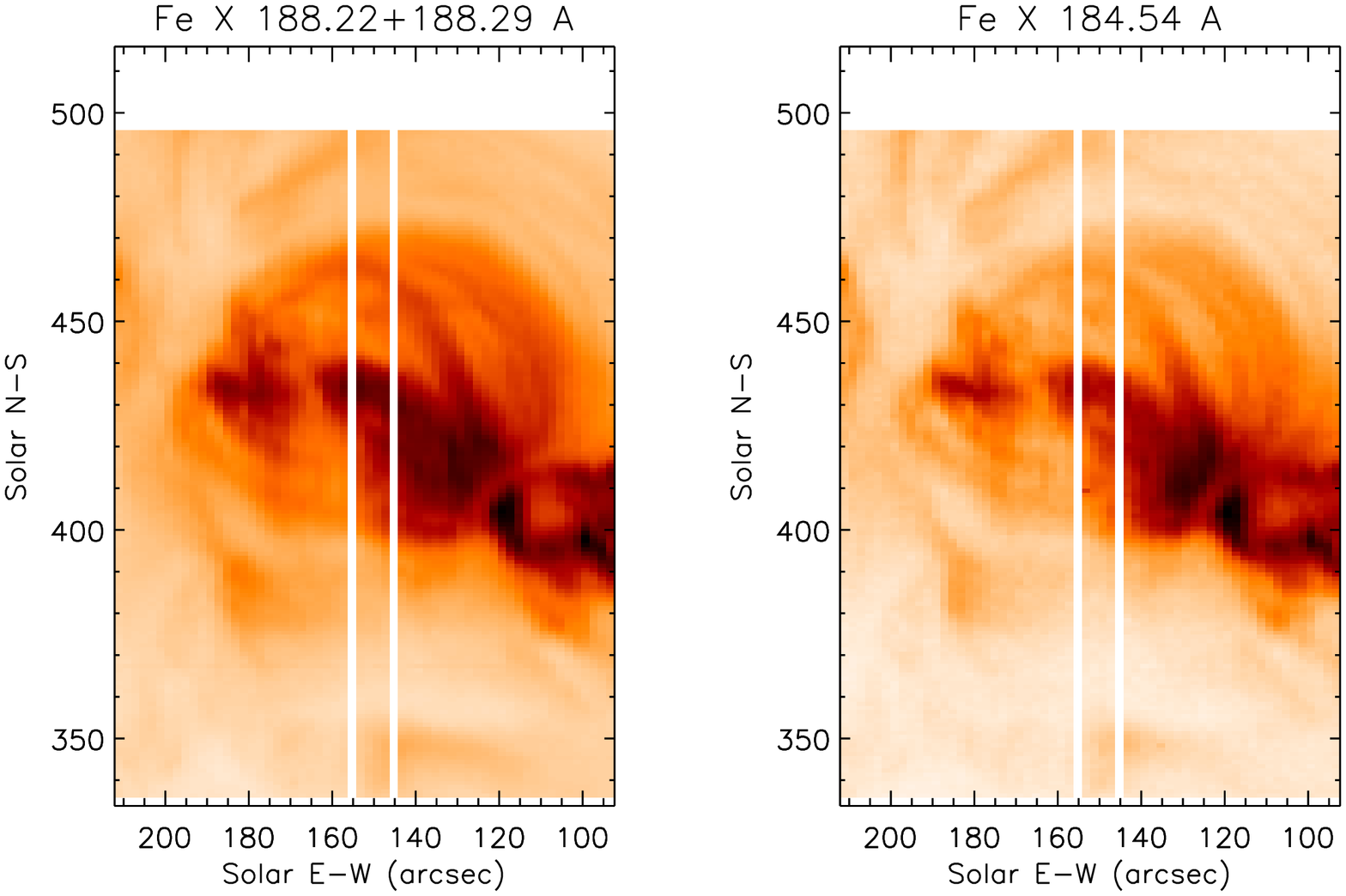}
\includegraphics[width=9.0cm,height=16.0cm,angle=90]{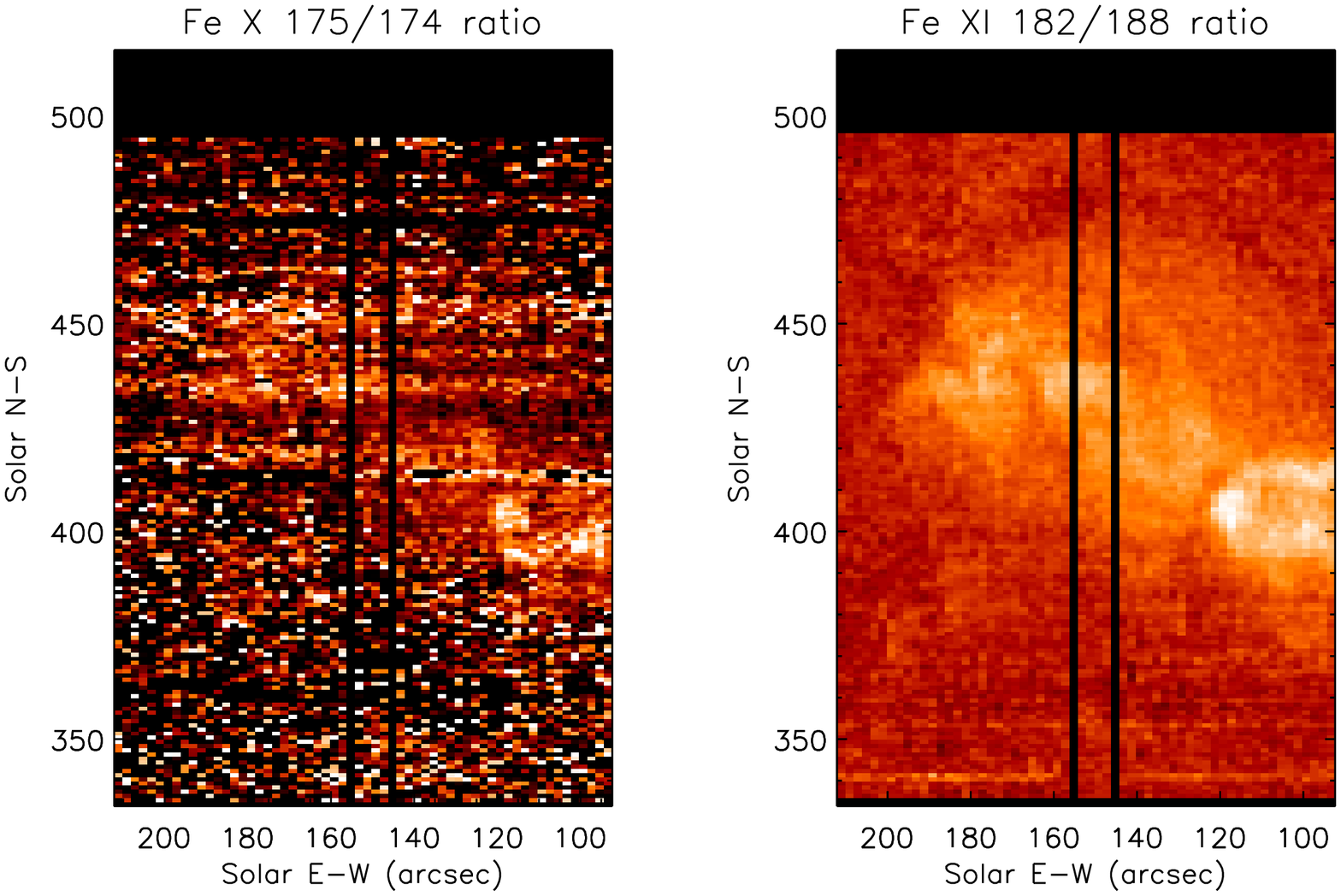}
\caption{EIS observations of AR11082 on June 2, 2007. {\bf Top:} intensity
maps obtained with the \ion[Fe xi] 188.2~\AA\ doublet and the \ion[Fe x]
184.54~\AA\ line. {\bf Bottom left:} Density map obtained with the \ion[Fe x]
175.26/174.54 intensity ratio; {\bf Bottom right:} Density map obtained with
the \ion[Fe xi] 182/188 intensity ratio.}
\label{dens2}
\end{figure}

However, using density measurements from another ion raises additional uncertainties: 
the emitting plasmas might not be the same, and thus the density measured by the two 
ions can be different. Figures~\ref{dens1} and \ref{dens2} show intensity maps of two 
active regions in the \ion[Fe x] 184.54~\AA\ line and \ion[Fe xi] 188.2~\AA\ doublet, 
showing that indeed the two ions sample the same plasmas. These observations were taken 
as examples because they also included the 174.54~\AA\ and 175.26~\AA\ lines, so that 
differences in the density values and their spatial distribution could be checked. Both 
active regions clearly show the difficulty in observing the two \ion[Fe x] 174.54~\AA\ 
and 175.26~\AA\ lines lines, due to the poor SNR, which is amplified in their ratio; 
still, the spatial distribution of the densest areas is the same as with the \ion[Fe xi] 
density ratio. A detailed comparison of density measurements over restricted areas show 
that these two ratios are within $\Delta \log N_e \approx 0.1$ in the brightest areas 
where the magnetic field measurements will be most accurate. Density values are in the 
$\log N_e=9.0-9.6$ range, where also the intensity ratios in Equations~\ref{mit} and 
\ref{m2} are density sensitive, so that electron density may indeed contribute to the
overall uncertainty in the measurement of the magnetic field strength.


\subsection{Intensity calibration}

Calibration is another source of uncertainty of critical importance. The reason is that
all the strongest \ion[Fe x] lines are located in the SW channel, while the magnetically
sensitive 257.26~\AA\ is located in the LW channel, so that the relative calibration of 
the two is uncertain. This uncertainty will directly propagate into the determination of 
both $I_{MIT}$ and $I_{M2}$ in Equations~\ref{mit} and \ref{m2}. Two independent studies
(HPW and GDZ) have been carried out to determine the in-flight EIS intensity calibration 
and compared it with the pre-launch one, from Lang \etal (2006).

Results showed that the pre-flight sensitivity of the LW channel was overestimated, but
GDZ and HPW disagreed on the amount. Even more importantly, the LW channel showed a 
degradation with time relative to the SW channel, which directly impacts the present 
diagnostic technique. The wavelength dependence of the effective areas within each 
channel was largely the same as the pre-flight calibration, with the exception of the 
shortest wavelength of the SW channel (GDZ) or both ends of the SW channel (HPW): in 
both cases, the use of the 174.53~\AA\ line for the measurement of the magnetic field 
proposed by Si \etal (2020b) is significantly affected. Also, contradictory results
were obtained when comparing the EIS absolute calibration with both the EVE instrument
on board SDO, and the rocket EUNIS flight (Wang \etal 2011). 

\begin{figure}[!t]
\includegraphics[width=5.0cm,height=5.0cm,angle=90]{final_20071210_001927_a.ps}
\includegraphics[width=5.0cm,height=5.0cm,angle=90]{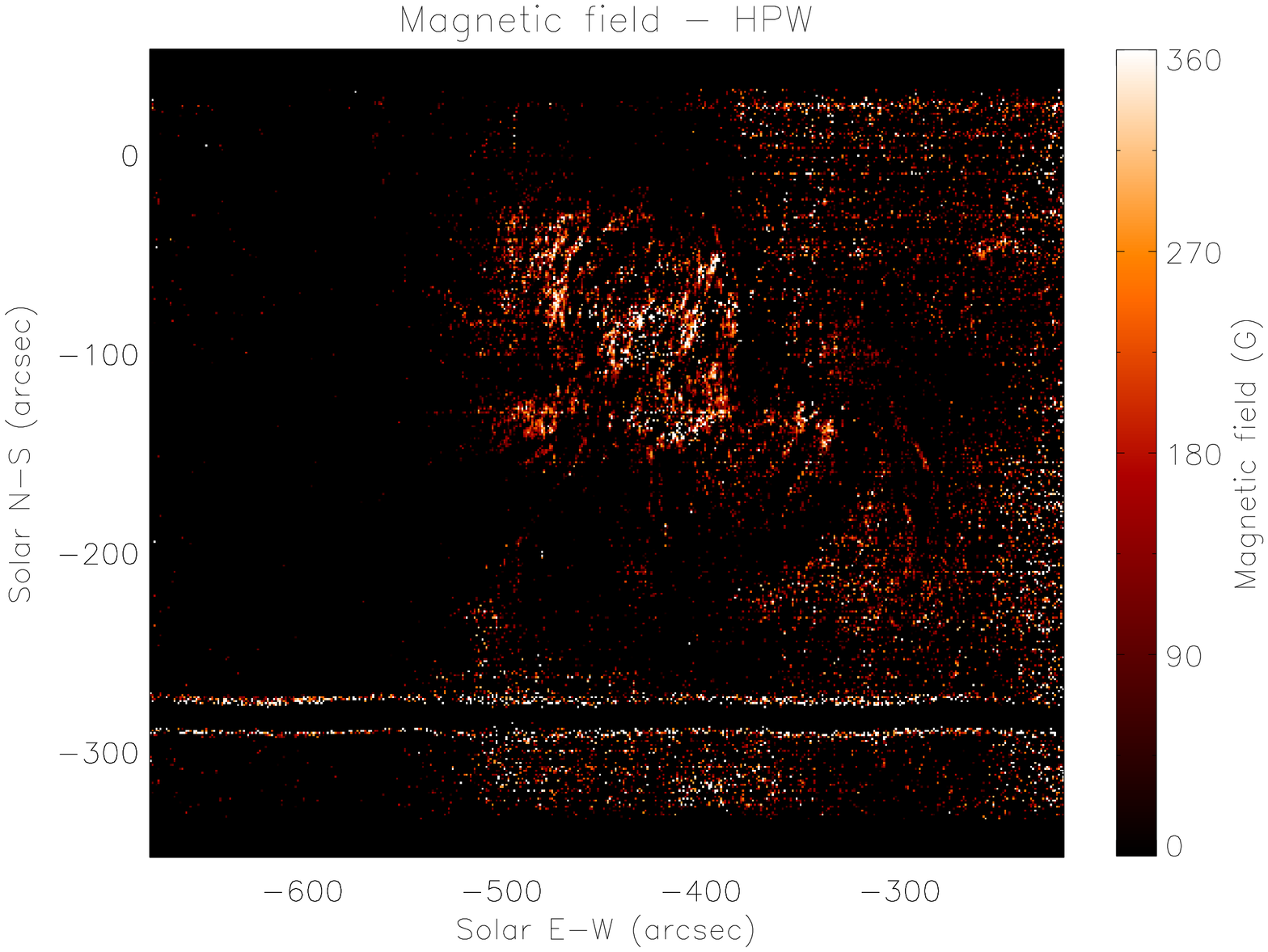}
\includegraphics[width=5.0cm,height=5.0cm,angle=90]{final_20071210_001927_d.ps}
\includegraphics[width=5.0cm,height=5.0cm,angle=90]{final_20071211_102542_a.ps}
\includegraphics[width=5.0cm,height=5.0cm,angle=90]{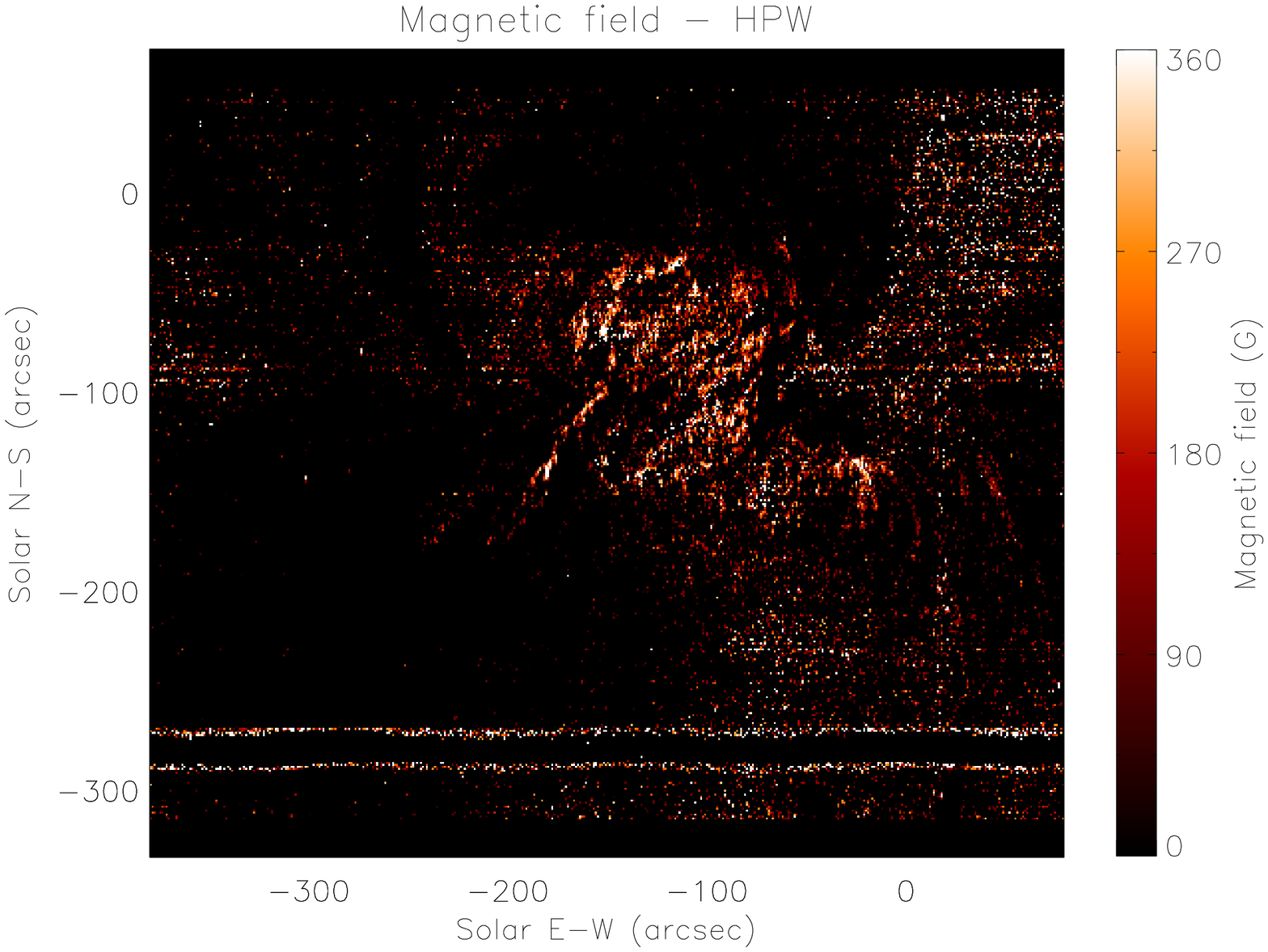}
\includegraphics[width=5.0cm,height=5.0cm,angle=90]{final_20071211_102542_d.ps}
\includegraphics[width=5.0cm,height=5.0cm,angle=90]{final_20071212_032643_a.ps}
\includegraphics[width=5.0cm,height=5.0cm,angle=90]{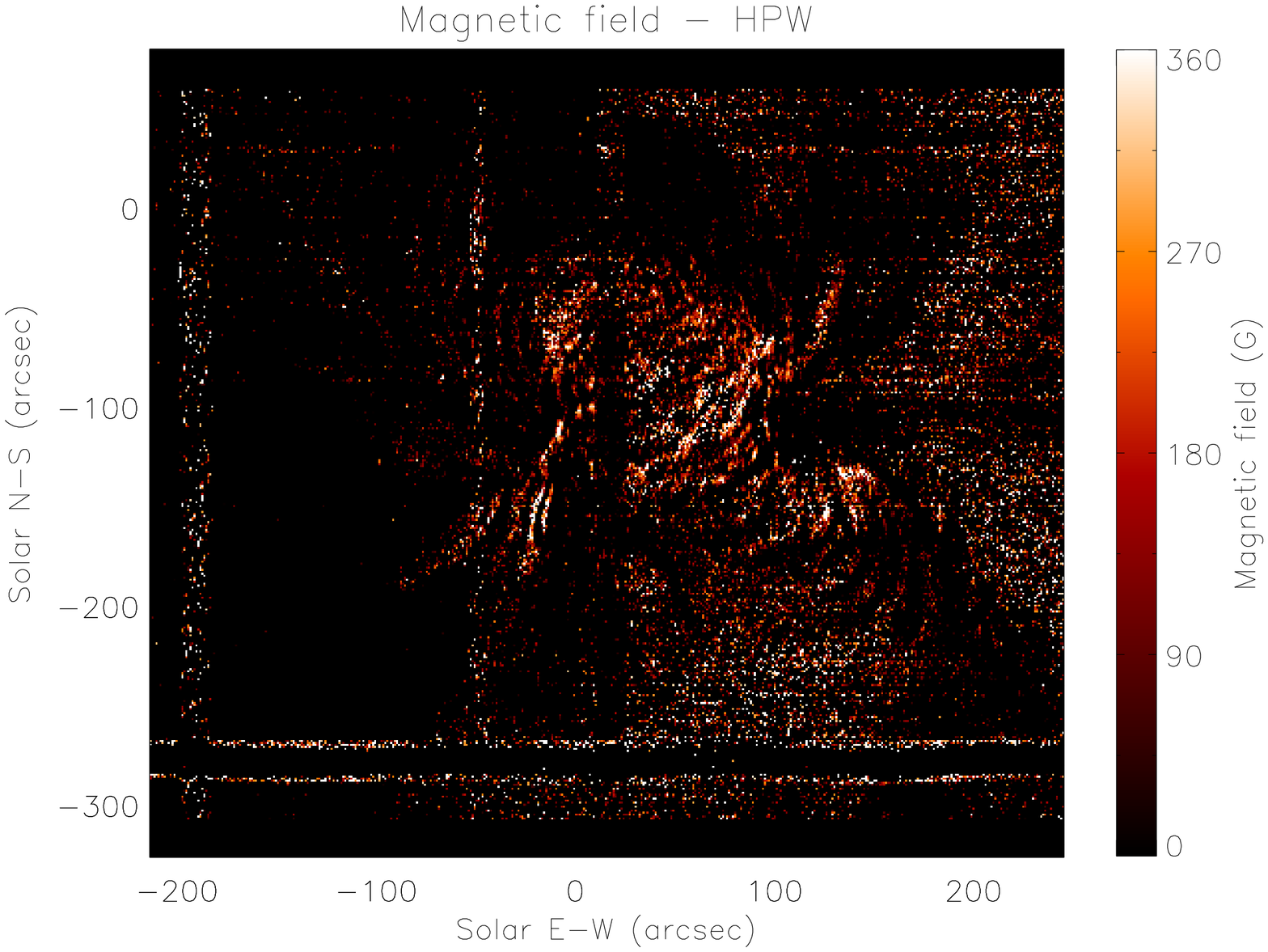}
\includegraphics[width=5.0cm,height=5.0cm,angle=90]{final_20071212_032643_d.ps}
\caption{Comparison of magnetic field diagnostic results obtained with the HPW calibration (middle panels)
and the GDZ calibration (right panels) from observations of AR10978.}
\label{cal1}
\end{figure}

\begin{figure}[!t]
\includegraphics[width=5.0cm,height=5.0cm,angle=90]{final_20071212_114336_a.ps}
\includegraphics[width=5.0cm,height=5.0cm,angle=90]{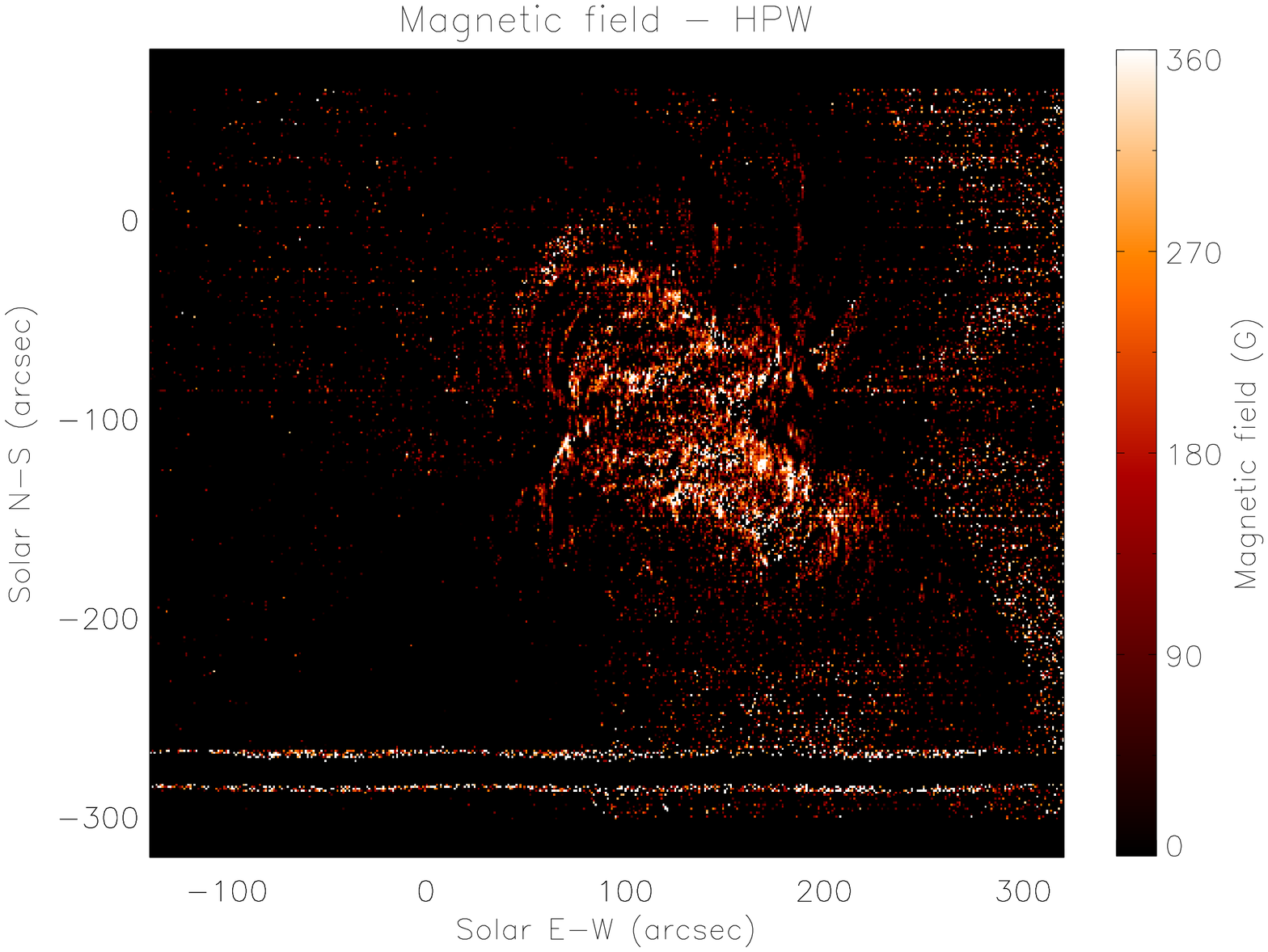}
\includegraphics[width=5.0cm,height=5.0cm,angle=90]{final_20071212_114336_d.ps}
\includegraphics[width=5.0cm,height=5.0cm,angle=90]{final_20071213_121842_a.ps}
\includegraphics[width=5.0cm,height=5.0cm,angle=90]{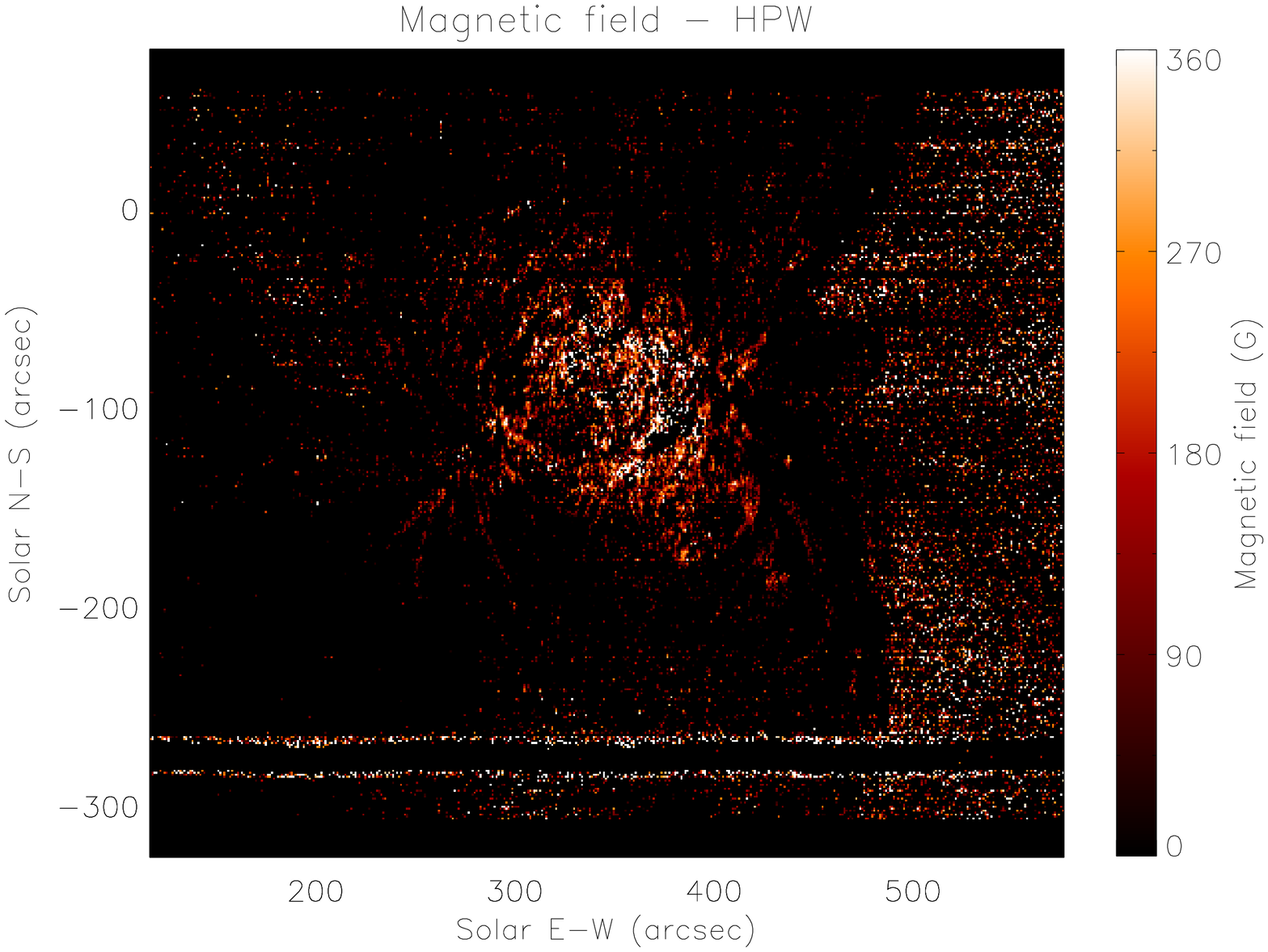}
\includegraphics[width=5.0cm,height=5.0cm,angle=90]{final_20071213_121842_d.ps}
\includegraphics[width=5.0cm,height=5.0cm,angle=90]{final_20071215_001349_a.ps}
\includegraphics[width=5.0cm,height=5.0cm,angle=90]{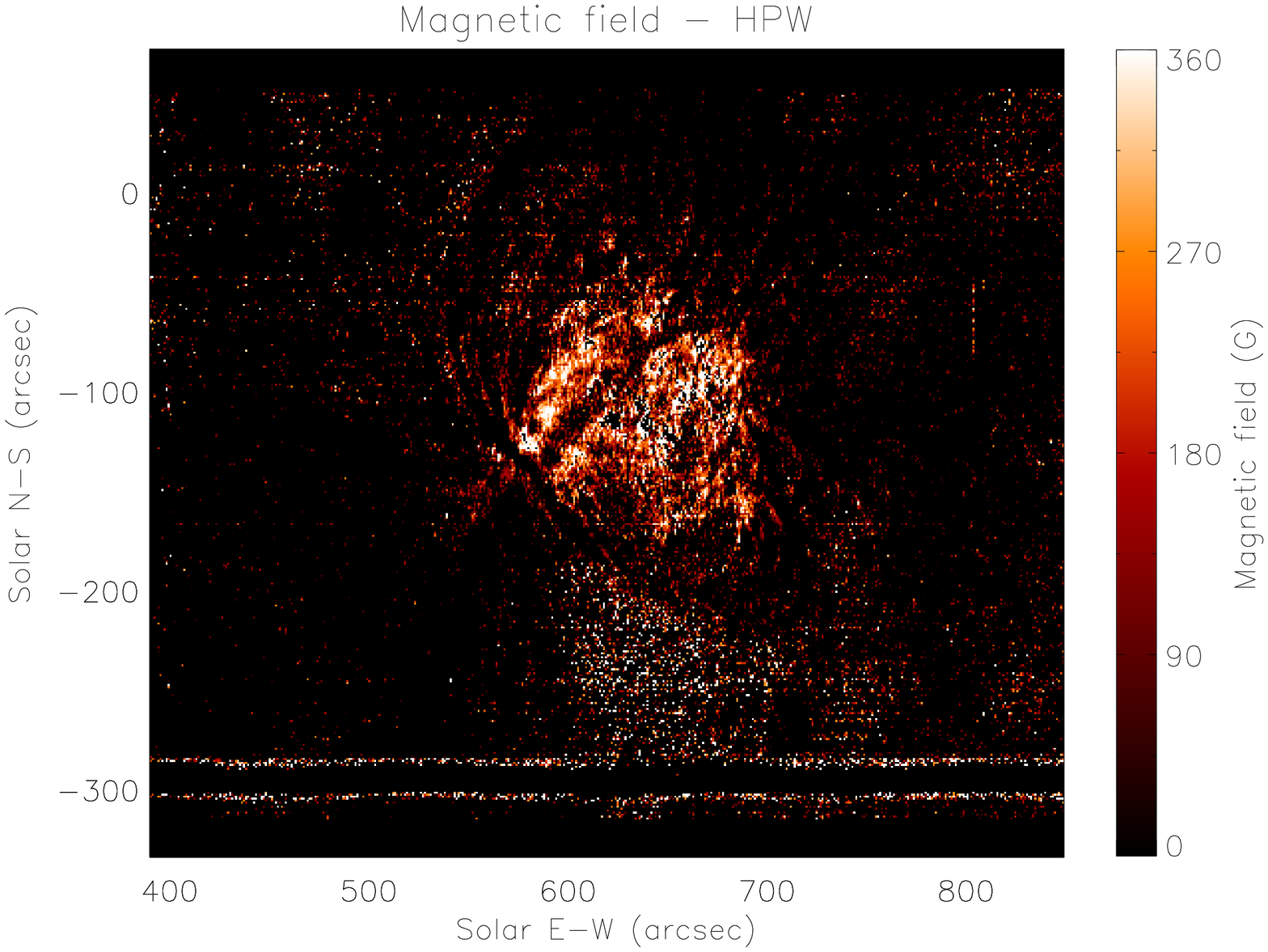}
\includegraphics[width=5.0cm,height=5.0cm,angle=90]{final_20071215_001349_d.ps}
\caption{Comparison of magnetic field diagnostic results obtained with the HPW calibration (middle panels)
and the GDZ calibration (right panels) from observations of AR10978.}
\label{cal2}
\end{figure}

These uncertainties and discrepancies directly affect the present diagnostic technique.
An example of calibration-related uncertainties is shown in Figures~\ref{cal1} and 
\ref{cal2}, which display some of the data shown in Figures~\ref{ar1} to \ref{ar3}. 
The left panels show \ion[Fe x] 184.54~\AA\ intensity maps, and the middle and right 
panels show magnetic field strength measurements obtained with HPW calibration (middle
panels) and GDZ calibration (right panels). As the scale of the magnetic field in both
panels is the same, differences are apparent, with the HPW calibration leading to higher
values for the stronger magnetic field, and giving zero magnetic field strength for 
regions where the GDZ calibration provides small, but measurable values of the magnetic
field. 
 
It is important to note that the uncertainties in the intensity calibration have a much
more limited impact on the determination of the plasma electron density, because this
parameter is measured using intensity ratios from lines close in wavelength, so that
any uncertainty is minimized. Nonetheless, we strongly urge the EIS instrument team to 
resume efforts to determine the EIS sensitivity for the entire duration of the mission.

\subsection{EIS detector sensitivity degradation}

\begin{figure}[!t]
\includegraphics[width=7.0cm,height=9.0cm,angle=90]{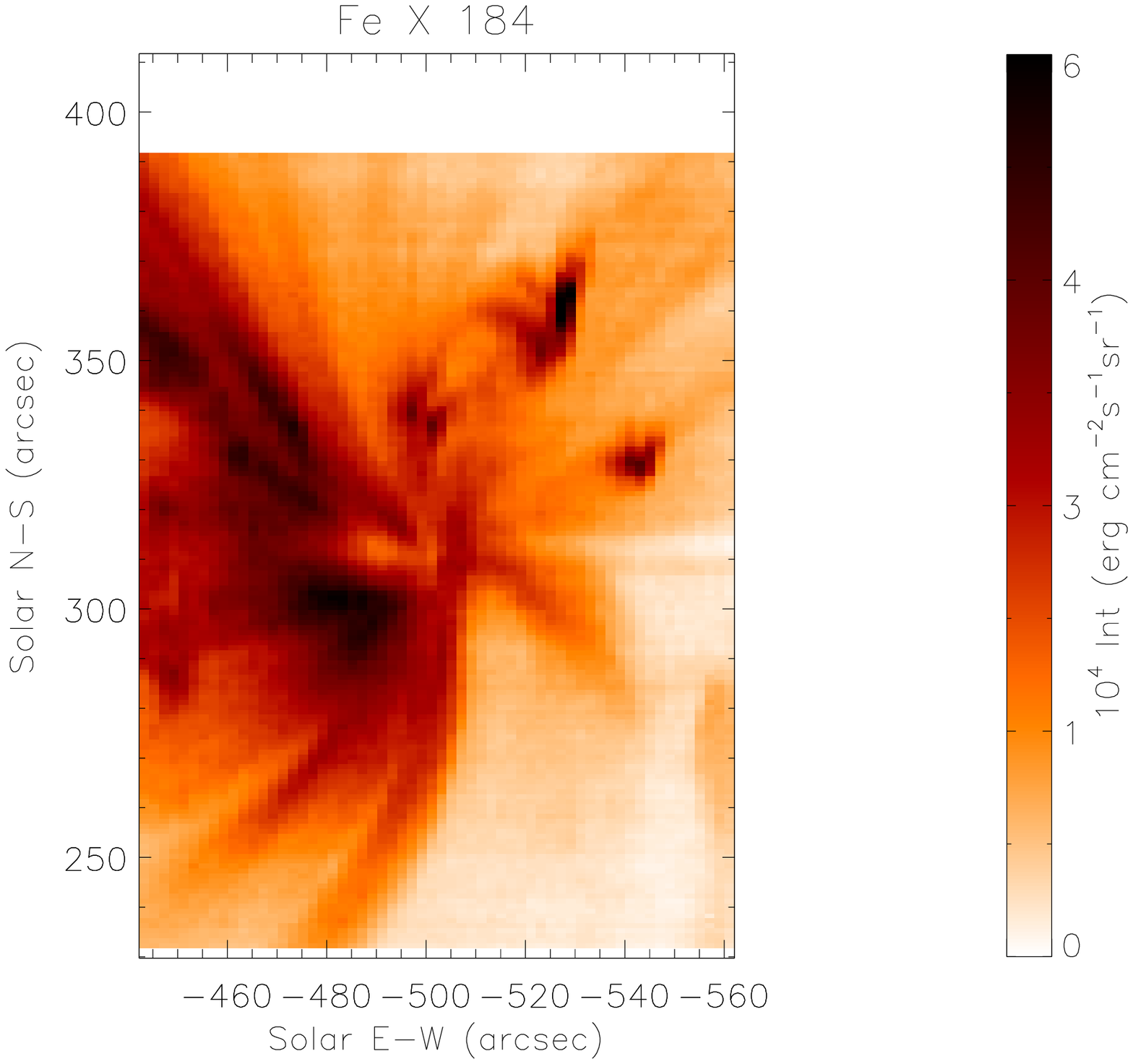}
\includegraphics[width=7.0cm,height=9.0cm,angle=90]{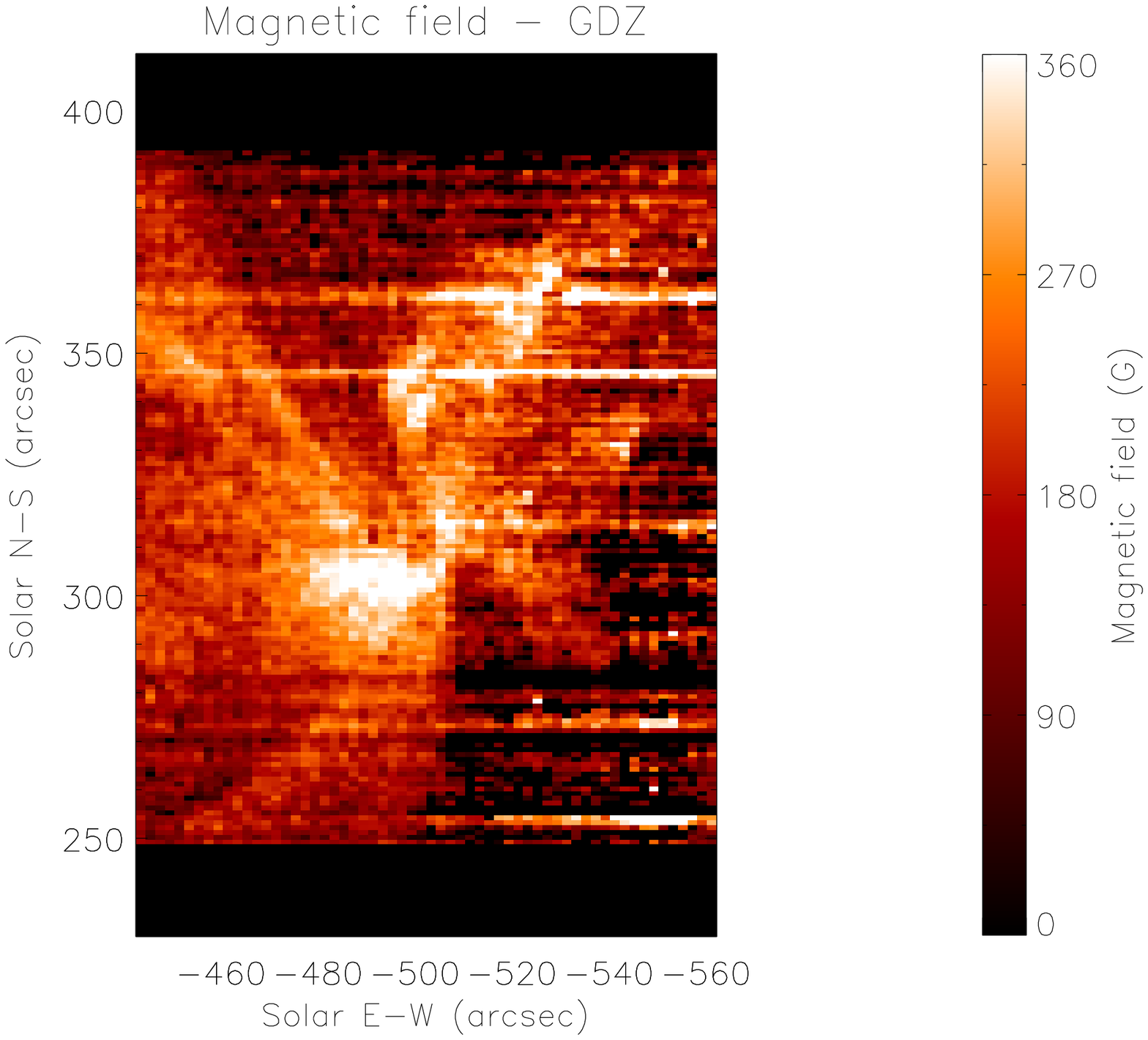}
\includegraphics[width=7.0cm,height=9.0cm,angle=90]{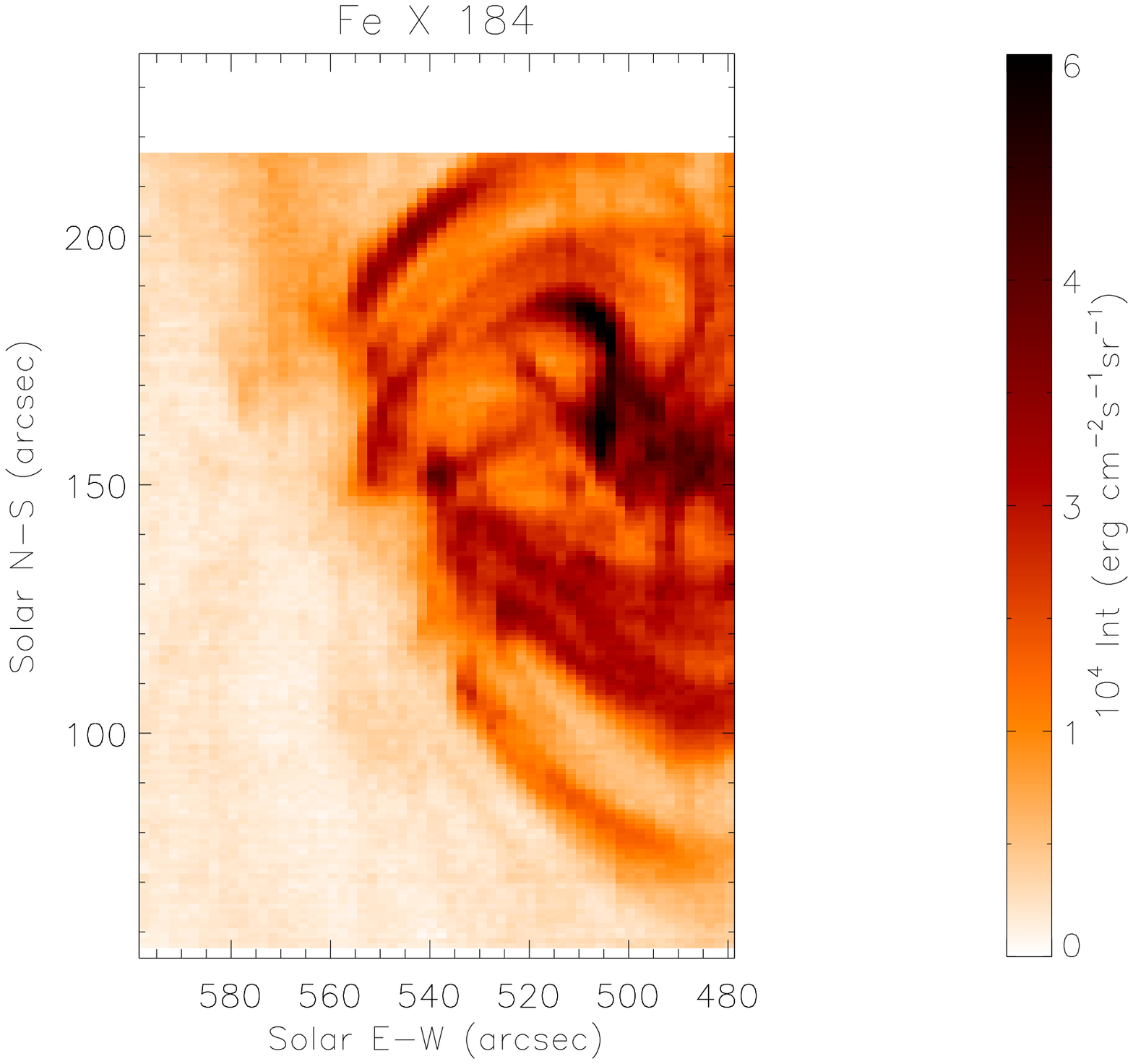}
\includegraphics[width=7.0cm,height=9.0cm,angle=90]{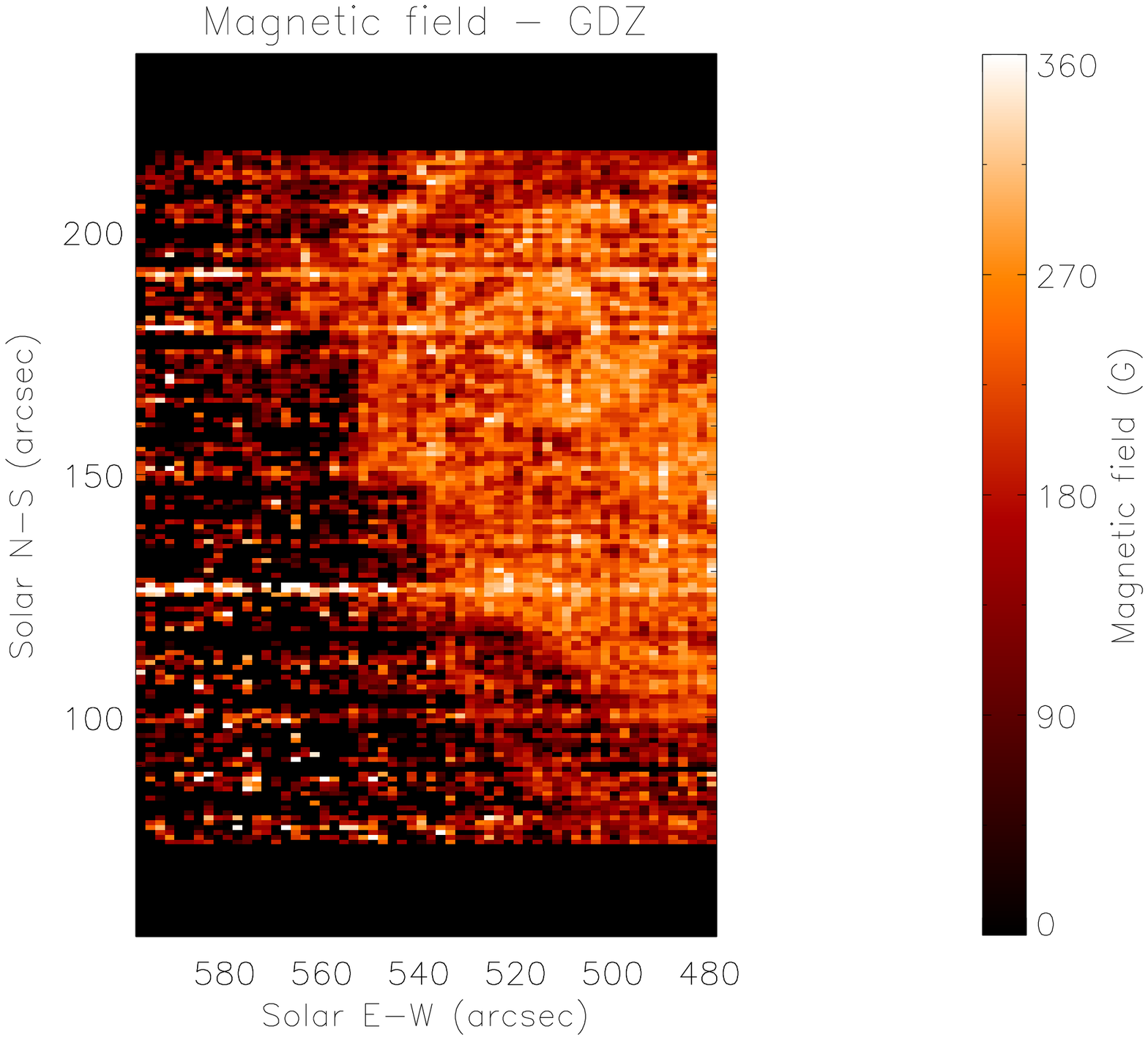}
\includegraphics[width=7.0cm,height=9.0cm,angle=90]{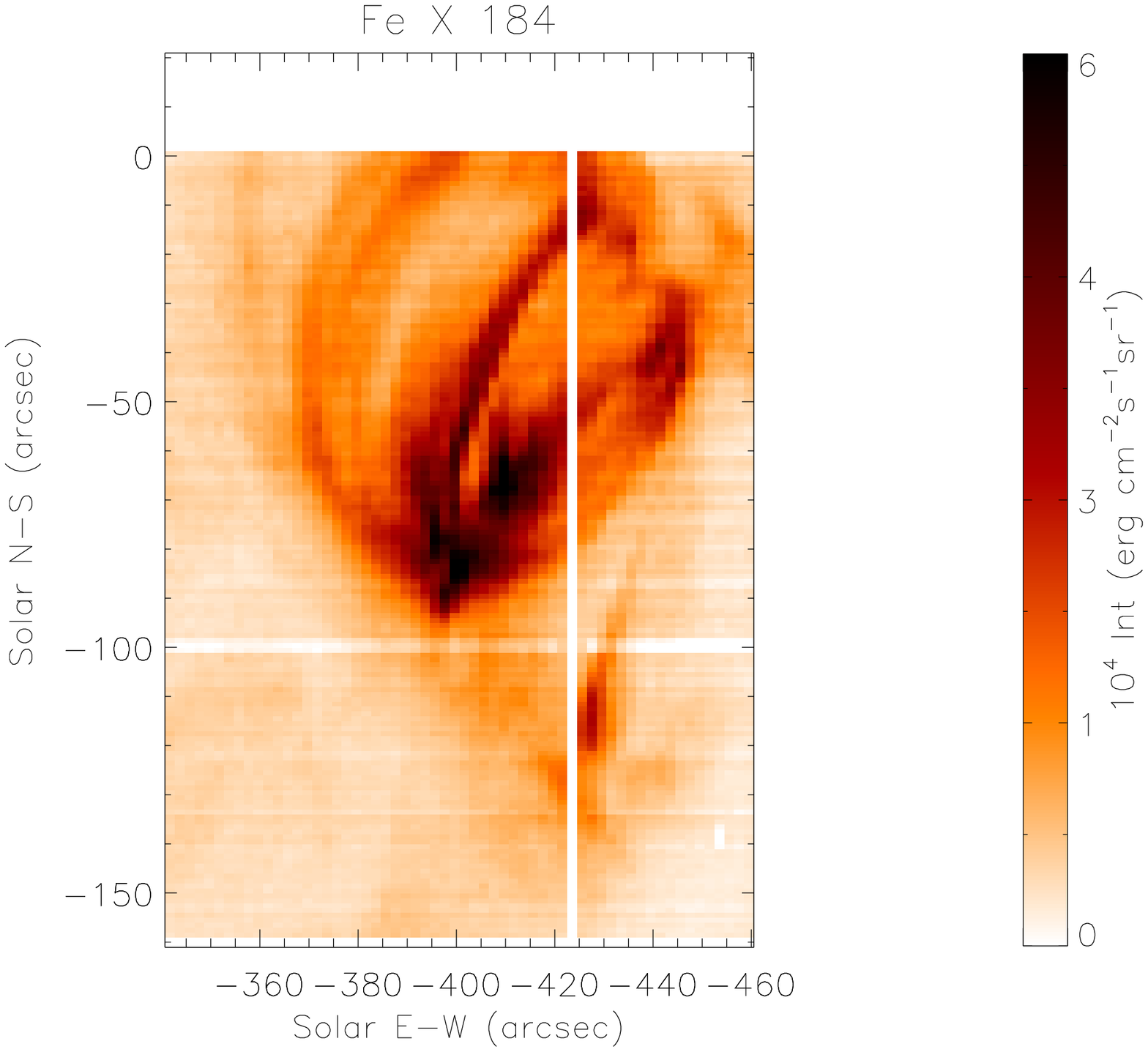}
\includegraphics[width=7.0cm,height=9.0cm,angle=90]{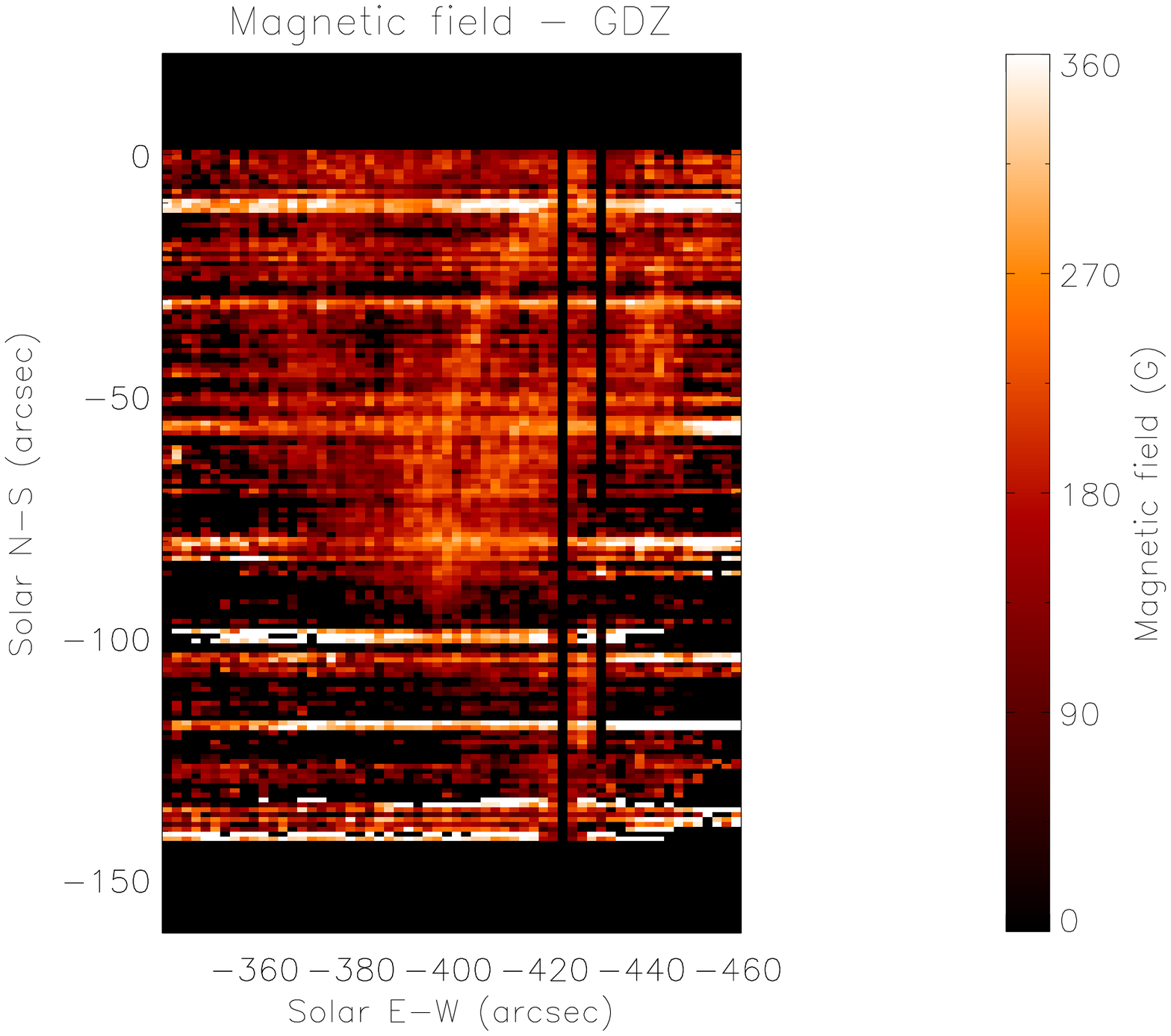}
\caption{Effects of the degradation of the EIS sensitivity. Left panels: \ion[Fe x] 184.53~\AA\
intensity map; right panels: magnetic field measurements. {\bf Top:} Active region observations 
taken on 16 March 2014 on AR12005; {\bf Middle:} active region observations taken on 25 October 
2016 on AR12603, {\bf Bottom:} Active region observations taken on 28 December 2018 on an
non-numbered AR.}
\label{degradation}
\end{figure}

Figure~\ref{degradation} shows the results of applying the magnetic field 
diagnostic technique to data sets taken in later stages of the EIS mission, where 
the instrument sensitivity has already significantly decreased, especially for the LW
channel where the magnetically sensitive 257.26~\AA\ line resides. Those results were 
obtained on full EIS spectra obtained with 60s exposure time and the 2" slit, a 
combination that at the beginning of the EIS mission ensured a very high SNR ratio.
While the intensity image has still high quality, the SNR ratio has degraded 
significantly, and the application of the diagnostic techniques, that needs a 
high SNR, gives results plagued by both instrumental effects and significantly 
more noise than earlier in the mission. Results can still be reliably obtained 
(especially considering that no rebinning or other correction was applied to 
these data sets shown in Figure~\ref{degradation}) but the degradation of the 
instrument is apparent.

\subsection{Atomic data}

The accuracy of the atomic data is also of critical importance for the present study. We have 
used CHIANTI version 9.0 to calculate line emissivities for all the levels of \ion[Fe x] 
as well as, when used for density diagnostics, \ion[Fe xi]. The original sources for these 
data are Del Zanna \etal (2010, 2013) for \ion[Fe xi], and Del Zanna \etal (2012) for 
\ion[Fe x]. In both cases large scale models for the atomic target were considered.
These works provided both energy levels, Einstein coefficients for spontaneous decay 
(A-values) and Maxwellian-averaged collision strengths.

The accuracy was benchmarked with observations from a number of laboratory, rocket, and 
space instrument measurements (Del Zanna 2011), as well as by Del Zanna \etal (2010, 2013) 
for \ion[Fe xi], Del Zanna \etal (2012) for \ion[Fe x], and more recently by Landi (2020, 
in preparation, both ions). These studies, using many observations from a number of 
different instruments, found that the intensities of the spectral lines used in this 
work for magnetic field diagnostics were in excellent agreement with observations. Most 
importantly, most of the observations used to test \ion[Fe x] were taken from quiescent 
spectra where the MIT transition is expected to be negligible, so that the ambient magnetic 
field did not affect the results.

Still, any uncertainty in the atomic data will directly affect magnetic field measurements.
In particular, at typical electron densities of active regions both the \tm[4 D 5/2,7/2] 
levels are populated mostly by radiative cascades from higher levels, so that the accuracy 
of both collision excitation rates and Einstein coefficients involving levels different 
than the \tm[4 D 5/2,7/2] is of great importance.

As an example, recently Wang \etal (2020) published a new, large scale calculation for 
energy levels and A-values of Cl-like ions which also included data for \ion[Fe x]. The 
much more extended model adopted in this calculation resulted in significant differences 
in the lifetimes and A values of many of the levels in the \orb[3s 2]\orb[3p 4]\orb[3d ] 
configuration; as the \tm[4 D 5/2,7/2] levels are mostly populated by cascades from 
\orb[3s 2]\orb[3p 4]\orb[3d ] levels with higher energy, these differences can have 
significant effects on their level population and therefore their line intensities. 
In order to assess the relevance of this effect, we have repeated all the measurements 
using an \ion[Fe x] model that combined the Wang \etal (2020) Einstein coefficients 
with the CHIANTI~9 collisional data, finding that the measured magnetic field strengths 
increased by 20-30\% with this hybrid model over the values obtained utilizing the 
CHIANTI~9 data for both collisional and radiative data. 

Unfortunately, Wang \etal (2020) only provided radiative data, lifetimes and energy levels,
but no collisional data, so that a self-consistent calculation of line intensities could
not be done. This comparison only underscore the need of new, large-scale calculations
of \ion[Fe x] collisional data using atomic models of the same accuracy as Wang \etal 
(2020). This is of course a formidable task, due to the added complexity in computing 
collision strengths, as well as other atomic properties involving continuum states, 
compared to rates in bound-bound transitions.

\section{Discussion and future work}
\label{discussion}

The present work illustrates the potential of a new diagnostic technique that allows the 
measurement of the magnetic field strength in active regions utilizing bright \ion[Fe x] 
and \ion[Fe xi] lines commonly observed by the Hinode/EIS satellite. This technique, 
which is based on a peculiar property found uniquely in two, near-degenerate \ion[Fe x] 
atomic levels, opens a new window on one of the most important, and less measured 
quantities in solar physics: the coronal magnetic field.

The potential of this new diagnostic technique lies in two basic facts: 1) that the coronal 
magnetic field determines all the critical processes at the heart of coronal heating, plasma 
confinement, and of the solar activity events (flares and CMEs) that give rise to Space Weather 
and to all the adverse effects it has on human assets on the ground and in space; 2) it can 
be applied to an existing data set spanning from 2007 to 2020 (as of this writing), extending
for more than one full solar cycle (including the anomalously weak minimum of solar cycle~24 
in 2007-2009), and covering a large number of active regions, flares, and even coronal mass
ejections. Thus, this new diagnostic technique opens the door for a whole new field of 
studies.

The importance of this technique is all the more enhanced by the development and deployment 
of the next generation ground based observatories which will be able to measure the coronal
magnetic field orientation and line-of-sight component at the solar limb: DKIST and UCoMP. When
combined together at the limb, EIS, DKIST and UCoMP observations will enable us to reconstruct
the full magnetic field in coronal active regions, providing for the first time an observable
of incalculable value for local and global models of active regions and of the solar corona.
Furthermore, unlike UCoMP and DKIST, EIS can measure the coronal magnetic field on the disk,
thus allowing the monitoring of active region fields as they transit across the disk: this 
capability enables the search of magnetic precursors to flares and CMEs, hopefully paving 
the way towards reliable Space Weather forecasting systems. Furthermore, this technique allows
to match magnetic field measurements to the determination of other properties that can be 
obtained with spectral lines from \ion[Fe x] and other elements with the same instrument,
providing a very complete characterization of active region plasmas.

In this work we present a few examples of the application of this technique to EIS observations
taken at different times, monitoring the morphology, strength and evolution of active region
magnetic fields. We find that the magnetic field strength in non-flaring active regions evolves
both in morphology and strength over time, and its value ranges from a few tens to a few hundred
Gauss. The present technique is alo able to provide 2D maps of the magnetic field in the plasma
that emit \ion[Fe x] line intensities.

Even more importantly, we present a list of EIS observing sequences whose data include all the 
lines necessary for the application of this diagnostic technique, and a few more that include 
density diagnostic line pairs from other ions that could also be used in some situations. We 
hope that this list will help the solar community navigate the immense set of EIS data and more 
easily find data to measure the coronal magnetic field.

However, this technique comes with limitations, that need to be overcome. The energy separation
between the \ion[Fe x] \tm[4 D 5/2,7/2] levels needs to be determined with even higher accuracy
than obtained by Landi \etal (2020), so that the intrinsic uncertainty on the $A_{MIT}$ value 
can be minimized. Also, the calibration of the EIS instrument needs to be determined both as 
a function of wavelength and as a function of time across the entire EIS mission, so that 
uncertainties related to the relative calibration within, and between, the SW and LW detectors 
can be minized as well. Also, we encourage the atomic physics community to improve on the 
radiative and collisional data for \ion[Fe x] over what is currently available, in order to 
minimize the effects of errors in the atomic and collision parameters on the measured magnetic 
field.

\acknowledgements

E. Landi was supported by NSF grants AGS 1408789, 1460170, and NASA grants NNX16AH01G, 
NNX17AD37G and 80NSSC18K0645. The authors would like to thank Dr. H.P. Warren for helpful 
discussions on the EIS intensity calibration. 

%\bibliography{cito}{}
%\bibliographystyle{aasjournal}

\end{document}